\documentclass[twocolumn,prd,aps]{revtex4}
\usepackage{graphicx,amsmath,amssymb,mathptmx}
\newcommand{\Lx}{\left(}
\newcommand{\Rx}{\right)}
\newcommand{\LB}{\left[}
\newcommand{\RB}{\right]}
\newcommand{\abbrev}[1]{\mbox{\small{#1}}}
\newcommand{\ep}{\epsilon}
\newcommand{\eqn}[1]{Eq.\,(\ref{#1})}

\newcommand{\reference}[1]{Ref.\,\cite{#1}}

\newcommand{\Section}[1]{Section\,(\ref{#1})}

\newcommand{\Appendix}[1]{Appendix\,(\ref{#1})}

\newcommand{\nlo}{\abbrev{NLO}}

\newcommand{\qcd}{\abbrev{QCD}}

\newcommand{\Rcal}{{\cal R}}

\newcommand{\bra}[1]{\left\langle #1\right|}
\newcommand{\ket}[1]{\left| #1\right\rangle}

\newcommand{\braket}[2]{\left.\VEV{#1\right| #2}}
\newcommand{\mtrxelm}[3]{\left\langle#1\left|#2\right|#3\right\rangle}
\newcommand{\ketbra}[2]{\left.\left.\left|#1\right\rangle\right\langle#2\right|}
\newcommand{\VEV}[1]{\left\langle #1\right\rangle}
\newcommand{\Tr}[1]{{\rm Tr}#1}

\newcommand{\overdot}[1]{\mathaccent"05F{#1}}
\newcommand{\spa}[2]{\left\langle#1#2\right\rangle}
\newcommand{\spb}[2]{\left[#1#2\right]}
\newcommand{\feynsl}[1]{{
  \setbox0=\hbox{/} \setbox1=\hbox{$#1$}
  \dimen0=\wd0 \advance\dimen0 by -\wd1 \divide\dimen0 by 2
  \ifdim\wd0>\wd1 \lower.15ex
          \copy0\kern-\wd0\kern\dimen0\copy1\kern\dimen0
  \else \kern-\dimen0\lower.15ex
          \copy0\kern-\dimen0\kern-\wd1\copy1\fi}}

\newcommand{\Inf}[2]{\LB{\mathop{\rm Inf}\nolimits}_{#1}\,#2\RB}
\newcommand{\Res}[3]{{\LB\frac{{\mathop{\rm Res}\nolimits}_{#1}\,#2}{#3}\RB}}
\begin{document}

\begin{titlepage}

\hspace*{\fill}\parbox[t]{2.8cm}{
\today
}

\bibliographystyle{apsrev}

\title{One-loop Integral Coefficients from Generalized Unitarity}

\author{William~B.~Kilgore}

\affiliation{Physics Department, Brookhaven National Laboratory,
  Upton, New York
  11973, U.S.A.\\
  {\tt [kilgore@bnl.gov]} }

\begin{abstract}
  I describe a method for determining the coefficients of
  scalar integrals for one-loop amplitudes in quantum field theory.
  The method is based upon generalized unitarity and the behavior of
  amplitudes when the free parameters of the cut momenta approach
  infinity.  The method works for arbitrary masses of both external
  and internal legs of the amplitudes.  It therefore applies not only
  to \qcd\ but also to the Electroweak theory and to quantum field
  theory in general.
\end{abstract}

\maketitle
\end{titlepage}


\section{Introduction}\label{sec::Intro}

For many years, high energy physics has been looking forward to the
beginning of the LHC program.  There are high expectations that
experiment will answer a number of outstanding theoretical issues,
such as the nature of electroweak symmetry breaking and the resolution
of the hierarchy problem.  However, the success of the experimental
program will depend upon the development of better theoretical tools.
Our ability to identify new physics at the TeV scale will depend, in
large part, on our ability to accurately describe and separate out the
known physics of the Standard Model.

It is well established that at the Tevatron and other colliders that
leading order calculations of \qcd\ processes are insufficient for an
accurate description of hadronic interactions.  Next-to-leading order
(\nlo) corrections in \qcd\ are often quite large.  The corrections
typically change both the predicted magnitude of the scattering
process and the shapes of distributions.

At the Tevatron, which has little reach into the multi-hundred GeV
range, it has been found that computing \nlo\ \qcd\ corrections is
generally sufficient to reduce the theoretical uncertainty of a
calculation to about 10\%.  At the LHC, however, with its reach
extending above the TeV scale, it has been found that electroweak
corrections can become quite large.  It is therefore becoming
important to develop tools for performing higher-order corrections in
the full $SU(3)\otimes SU(2)\otimes U(1)$ Standard Model.

In the past, many of the most exotic computational methods focused on
\qcd\, not just because \qcd\ corrections are large even at relatively low
energies, but also because working with massless quarks and gluons
simplifies expressions sufficiently to allow complicated algorithms to
be worked out.  Recent advances now provide a framework in which
massive theories can access the sophisticated advances developed for
\qcd\ calculations.

The workhorse in perturbative calculations has long been the Feynman
diagram approach, which approach systematically includes all
perturbative effects.  However, the rapid growth in the number of
Feynman diagrams with the number of external legs, even for tree-level
amplitudes, limits the practical application of this technique.  In
addition, Feynman diagram calculations are subject to large
cancellations among individual diagrams, especially when working with
gauge theories.

One tool that has long been used in \qcd\ calculations to obtain
compact results and largely avoid unphysical singularities and gauge
cancellations is the helicity
method~\cite{Berends:1981rb,DeCausmaecker:1981bg,Xu:1984qe,
Kleiss:1985yh,Gunion:1985vca,Hagiwara:1985yu,Xu:1986xb,Mangano:1991by}.
When combined with recursion relations, both
off-shell~\cite{Berends:1987me,Kosower:1989xy} and, more recently,
on-shell~\cite{Britto:2004ap,Britto:2005fq,Bern:2005hs,Bern:2005ji}
the helicity method can be used to generate compact expressions for
for very complicated scattering processes.

Another very powerful technique that has greatly simplified the
calculation of loop amplitudes in \qcd\ is the unitarity
method~\cite{Bern:1994zx,Bern:1994cg,
Bern:1995db,Bern:1996ja,Bern:1996je,Bern:1996fj,Bern:2000dn,Bern:2002tk}.
An essential feature of the unitarity method is that it sews together
tree-level amplitudes into loop amplitudes.  Thus, efficient
techniques for computing tree-level amplitudes which give compact
expressions, like the use of recursion relations and the helicity
method, directly benefit the unitarity method of computing loop
amplitudes.

There have been a number of new developments that have
significantly enhanced the power of recursive methods and the
unitarity method.  These include the use of maximally
helicity-violating (MHV) vertices in recursion relations at
tree-level~\cite{Cachazo:2004kj,Zhu:2004kr,Georgiou:2004wu,Wu:2004fba,
Kosower:2004yz,Georgiou:2004by,Abe:2004ep,Dixon:2004za,Bern:2004ba,
Birthwright:2005ak,Risager:2005vk} and in
loops~\cite{Brandhuber:2004yw,Quigley:2004pw,Bedford:2004py,Bedford:2004nh}.
Unitarity methods have been improved by the use of the holomorphic
anomaly~\cite{Cachazo:2004by,Cachazo:2004dr,Britto:2004nj} to evaluate
cuts and the use of complex momenta~\cite{Britto:2004nc,
Brandhuber:2005jw,Quigley:2005cu,Britto:2005ha,Britto:2006sj} within
the context of generalized unitarity~\cite{Mandelstam:1958xc,
Landau:1959fi,Cutkosky:1960sp,Eden:1966bk,Bern:1997sc,Bern:2004cz,Bern:2004ky},
which allows for the use of multiple cuts in a single amplitude.

These techniques can be combined into a ``unitarity
bootstrap''~\cite{Bern:2005cq,Berger:2006ci}, a systematic, recursive 
approach to making high-multiplicity \qcd\ calculations practical.
The bootstrap combines the use of four-dimensional unitarity to
determine the logarithmic and polylogarithmic terms in the loop
amplitude with on-shell recursion relations to determine the rational
contributions.

Recently, Forde~\cite{Forde:2007mi} has described an efficient method
for extracting the coefficients of the triangle and bubble functions
in amplitudes where only massless particles circulate in the loops.
Combined with the result of \reference{Britto:2004nc}, this is
sufficient to determine the cut-constructable part of one-loop QCD
amplitudes.  In this paper, I will extend Forde's formalism and define
methods to extract the coefficients of all loop-integral functions
in massive as well as massless theories.  This will then allow this
method to be used in the full $SU(3)\otimes SU(2)\otimes U(1)$
Standard Model and even in the Supersymmetric Standard Model.

The plan of the paper is as follows: In \Section{sec::notation}, I
will discuss the notation used and the application of spinor helicity
conventions to massive complex momenta.  In \Section{sec::methods}, I
will present an overview of the formalism which allows the simple
extraction of the loop integral functions.  In \Section{sec::box}, I
will discuss the extraction of the box functions; in
\Section{sec::triangle} the extraction of the triangle functions; in
\Section{sec::bubble}, the bubble functions; and in
\Section{sec::tadpole}, the tadpoles.  Finally, I will comment on my
results and draw my conclusions.


\section{Notation and Conventions}
\label{sec::notation}
An important feature of the method described in this paper is that it
involves on-shell amplitudes where some of the on-shell momenta are
complex.  Therefore, I will describe the conventions used to handle
complex momenta and, especially, the spinorial representations of
complex momenta.

\subsection{Spinor Representations of Real Momenta}
\label{sec::massless}
I treat real momenta using the spinor-helicity formalism of Mangano
and Parke~\cite{Mangano:1991by}, which I will summarize briefly.  Let
$p^\mu$ be a massless momentum in Minkowski space, and $\psi(p)$
the Dirac spinor representing a massless fermion of momentum $p$,
\begin{equation}
\feynsl{p}\,\psi(p) = 0,\qquad p^2 = 0\,.
\end{equation}
The two helicity states of $\psi(p)$ are given by
\begin{equation}
\begin{split}
\psi_{\pm}(p) &=\ \frac{1}{2}\Lx1\pm\gamma_5\Rx\psi(p)
   \equiv\ \ket{p^\pm}\,,\\
\overline{\psi}_{\pm}(p) &=\
   \frac{1}{2}\overline{\psi}(p)\Lx1\pm\gamma_5\Rx
   \equiv\ \bra{p^\pm}\,.
\end{split}
\end{equation}
The phase convention is chosen such that
\begin{equation}
\label{eq::phases}
\ket{p^\pm} = \ket{p^\mp}^c\,,\qquad
   \bra{p^\pm} = {}^c\bra{p^\mp}\,,
\end{equation}
where the $c$ indicates charge conjugation.  To render expressions
more compact, I adopt the notation
\begin{equation}
\begin{split}
\spa{p}{q} &=\ \braket{p^-}{q^+}\,,\\
\spb{p}{q} &=\ \braket{p^+}{q^-}\,.
\end{split}
\end{equation}
Some important identities are:
\begin{equation}
\begin{split}
\braket{p^+}{q^+} &=\ \braket{p^-}{q^-} =\ 0\,,\\
\spa{p}{p} &=\ \spb{p}{p} =\ 0\,,\\
\spa{p}{q} &=\ -\spa{q}{p}\qquad \spb{p}{q} =\ -\spb{q}{p}\,,\\
\spa{p}{q}\spb{q}{p} &=\ 2\,p\cdot q\,,\\
\mtrxelm{p^\pm}{\gamma^\mu}{p^\pm} &=\ 2\,p^\mu\,,\\
\feynsl{p} &=\ \ketbra{p^+}{p^+} + \ketbra{p^-}{p^-}\,,\\
\gamma_\mu\mtrxelm{p^\pm}{\gamma^\mu}{q^\pm} &=\ 
   2\Lx\ketbra{p^\mp}{q^\mp} + \ketbra{q^\pm}{p^\pm}\Rx\,,\\
\spa{A}{B}\spa{C}{D} &=\ \spa{A}{C}\spa{B}{D} + \spa{A}{D}\spa{C}{B}\,,\\
\spb{A}{B}\spb{C}{D} &=\ \spb{A}{C}\spb{B}{D} + \spb{A}{D}\spb{C}{B}\,.
\end{split}
\end{equation}
The last two identities are known as the Schouten identities.

This formalism can alternatively be expressed in terms of Weyl-van der
Waarden (WvdW) spinors~\cite{Dittmaier:1998nn}.  Indeed there is a one-to-one
correspondence between the spinors defined here and WvdW spinors:
\begin{equation}
\begin{split}
\ket{\phi^+}\ \leftrightarrow\ \genfrac{(}{)}{0pt}{}{\phi_A}{0}\,,&\qquad
  \bra{\phi^-}\ \leftrightarrow\ \Lx\phi^A\quad 0\Rx\,,\\
\ket{\chi^-}\ \leftrightarrow\ \genfrac{(}{)}{0pt}{}{0}{\chi^{\overdot{A}}}\,,&\qquad
  \bra{\chi^+}\ \leftrightarrow\ \Lx0\ \quad\chi_{\overdot{A}}\Rx\,,\\
\spa{\phi}{\chi}_{MP}\leftrightarrow\spa{\phi}{\chi}_{WvdW}\,,&\qquad
    \spb{\chi}{\phi}_{MP}\leftrightarrow\spa{\phi}{\chi}^*_{WvdW}\,,\\
   \qquad\gamma^\mu \leftrightarrow&
   \genfrac{(}{)}{0pt}{}{0\quad\qquad\sigma^{\mu}_{A\overdot{B}}}
   { \sigma^{\mu,\,A\overdot{B}}\qquad 0\quad}\,,
\end{split}
\end{equation}
Where the $MP$ and $WvdW$ subscripts denote the ``Mangano \& Parke''
and ``Weyl-van der Waerden'' notation for spinor products,
respectively.

\subsubsection{Fermion Wave Functions for Real Massless Momenta}
\label{sec::masslessferm}
These are essentially trivial, since the notation is defined in terms
of massless Dirac spinors.  The only point of convention is that, for
the sake of defining helicity, all external particles are taken to be
outgoing.  This means that we identify
\begin{equation}
\ket{p^\pm} =\ v_\pm(p)\,,\qquad \bra{p^\pm} =\ \overline{u}_\pm(p)\,.
\end{equation}
The phase convention defined in \eqn{eq::phases} means that
\begin{equation}
u_\pm(p) = v_\mp(p)\,,\qquad \overline{u}_\pm(p) =
  \overline{v}_\mp(p)\,.
\end{equation}

\subsubsection{Massless Vector Boson Wave-functions for Real Momenta}
Massless spin-1 particles have two physical polarization states.  The
standard practice in spinor helicity methods is to use the light-like
axial gauge, in which the polarization vectors are defined in terms of
both the momentum vector $k$ and a reference vector $g$.  The gauge
invariance of the spin-1 field manifests itself in the arbitrariness
of the reference momentum.  For an outgoing vector field of
momentum $k$, the helicity states are can be written as
\begin{equation}
\varepsilon^{+\,\mu}(k;g) = \frac{\mtrxelm{k^+}{\gamma^\mu}{g^+}}
     {\sqrt{2}\spa{g}{k}}\,,\quad
    \varepsilon^{-\,\mu}(k;g) = \frac{\mtrxelm{k^-}{\gamma^\mu}{g^-}}
     {\sqrt{2}\spb{k}{g}}\,.
\end{equation}
These polarization vectors have the usual properties
\begin{equation}
\begin{split}
\Lx\varepsilon^\pm\Rx^* &=\ \varepsilon^\mp\,,\\
\varepsilon^\pm\cdot\varepsilon^\pm &=\ 0\,,\\
\varepsilon^\pm\cdot\varepsilon^\mp &=\ -1\,,\\
\varepsilon^{+\,\mu}\varepsilon^{-\,\nu} &+
   \varepsilon^{-\,\mu}\varepsilon^{+\,\nu} =
   - g^{\mu\nu} + \frac{k^\mu\,g^\nu + g^\mu\,k^\nu}{g\cdot k}\\
\end{split}
\end{equation}
The arbitrariness of the choice of $g$ can be seen by examining the
difference between two choices of $g$.  Consider contracting a
polarization vector with some random vector,
\begin{equation}
\begin{split}
\Lx\varepsilon^{+\,\mu}(k;g_1) \right.&\left.- \varepsilon^{+\,\mu}(k;g_2)\Rx
    \mtrxelm{a^-}{\gamma_\mu}{b^-}\\
  &=\ \sqrt{2}\Lx\frac{\spb{k}{b}\spa{a}{g_1}}{\spa{g_1}{k}}
     - \frac{\spb{k}{b}\spa{a}{g_2}}{\spa{g_2}{k}}\Rx\\
  &=\ \sqrt{2}\spb{k}{b}\frac{\spa{a}{g_1}\spa{g_2}{k} -
   \spa{a}{g_2}\spa{g_1}{k}}{\spa{g_1}{k}\spa{g_2}{k}}\\
  &=\ \sqrt{2}\spb{k}{b}\frac{\spa{a}{k}\spa{g_2}{g_1}}
   {\spa{g_1}{k}\spa{g_2}{k}}\\
  &=\ \frac{k^\mu\,\spa{g_1}{g_2}}{\sqrt{2}\spa{g_1}{k}\spa{g_2}{k}}
   \mtrxelm{a^-}{\gamma_\mu}{b^-}\,.
\end{split}
\end{equation}
Thus, the difference between the polarization vectors generated by two
choices of $g$ is proportional to $k^\mu$.  Since gauge bosons couple
to conserved currents, this is a pure gauge term and does not
contribute to the amplitude.

\subsection{Massive Real Momenta}
A massive real momentum can be represented as the sum of two massless
momenta.  There is great freedom in choosing this decomposition, which
leads to a variety of choices for representing massive fermion and
vector wave functions.  The construction of helicity basis wave
functions for massive particles is discussed thoroughly in
\reference{Dittmaier:1998nn} and is not repeated here.  One can even
choose the decomposition so that the massive wave functions are
eigenstates of the helicity projection operator.

\subsection{Massless Complex Momenta}
The massless complex momenta that I will encounter will be defined in
terms of the spinor representations of two real massless momenta
$\chi^\mu$ and $\psi^\mu$,
\begin{equation}
\begin{split}
\ell^\mu &=\ y\,\chi^\mu + w\,\psi^\mu +
\frac{t}{2}\mtrxelm{\chi^-}{\gamma^\mu}{\psi^-}+ \frac{w y}{2t}\mtrxelm{\psi^-}{\gamma^\mu}{\chi^-}.
\end{split}
\end{equation}
One can define a spinor representation of this complex momentum as
\begin{equation}
\begin{split}
\bra{\ell^+} =\ \frac{y}{t}\bra{\chi^+} + \bra{\psi^+}\,,
      \qquad& \bra{\ell^-} =\ t\bra{\chi^-} + w\bra{\psi^-}\,,\\
\ket{\ell^-} =\ \frac{y}{t}\ket{\chi^-} + \ket{\psi^-}\,,
      \qquad&\ket{\ell^+} =\ t\ket{\chi^+} + w\ket{\psi^+}\,.\\
\end{split}
\end{equation}
Note that the different helicity states are not related by complex
conjugation, as they are in the case of real momenta.  As in the case
of real momenta, however, these spinor representations can be used
directly as massless fermion wave function and in massless vector
wave-functions (see \Section{sec::massless}).

\subsection{Massive Complex Momenta}
Let us assume that we have an on-shell complex momentum $\ell^\mu$
with mass $m$ parametrized in terms of real massless four-momenta
$\chi^\mu$ and $\psi^\mu$,
\begin{equation}
\begin{split}
\ell^\mu &=\ y\,\chi^\mu + w\,\psi^\mu +
\frac{t}{2}\mtrxelm{\chi^-}{\gamma^\mu}{\psi^-}\\
&\hskip 40pt+ \Lx\frac{w y}{2t} - \frac{m^2}{4t \chi\cdot\psi}\Rx
  \mtrxelm{\psi^-}{\gamma^\mu}{\chi^-}.
\end{split}
\end{equation}
I can trivially decompose $\ell^\mu$ into two massless complex
momenta,
\begin{equation}
\begin{split}
\ell^\mu &=\ \ell_1^\mu + \ell_2^\mu\ ,\\
\ell_1^\mu &=\ y\,\chi^\mu + w\,\psi^\mu +
\frac{t}{2}\mtrxelm{\chi^-}{\gamma^\mu}{\psi^-} + \frac{w y}{2t}
  \mtrxelm{\psi^-}{\gamma^\mu}{\chi^-}\ ,\\
\ell_2^\mu &=\ - \frac{m^2}{4t \chi\cdot\psi}\mtrxelm{\psi^-}{\gamma^\mu}{\chi^-},
\end{split}
\end{equation}
and then find spinor representations of $\ell_1^\mu$ and $\ell_2^\mu$,
\begin{equation}
\begin{split}
\label{eq::spinorrep}
\bra{\ell_1^+} =\ \frac{y}{t}\bra{\chi^+} + \bra{\psi^+}
      \qquad& \bra{\ell_1^-} =\ t\bra{\chi^-} + w\bra{\psi^-}\\
\ket{\ell_1^-} =\ \frac{y}{t}\ket{\chi^-} + \ket{\psi^-}
      \qquad&\ket{\ell_1^+} =\ t\ket{\chi^+} + w\ket{\psi^+}\\
\bra{\ell_2^+} =\ \frac{m}{\spb{\chi}{\psi}}\bra{\chi^+}\hskip38pt&
       \bra{\ell_2^-}=\ \frac{m}{t\,\spa{\chi}{\psi}}\bra{\psi^-}\\
\ket{\ell_2^-} =\ \frac{m}{\spb{\chi}{\psi}}\ket{\chi^-}\hskip38pt&
       \ket{\ell_2^+}=\ \frac{m}{t\,\spa{\chi}{\psi}}\ket{\psi^+},\\
\feynsl{\ell} =\ \ketbra{\ell_1^+}{\ell_1^+} + \ketbra{\ell_1^-}{\ell_1^-}
       +& \ketbra{\ell_2^+}{\ell_2^+} + \ketbra{\ell_2^-}{\ell_2^-}.
\end{split}
\end{equation}

With this parametrization, one immediately finds that
\begin{equation}
\label{eq::ell1ell2}
\spa{\ell_1}{\ell_2} = -\spa{\ell_2}{\ell_1} = m, \qquad
\spb{\ell_2}{\ell_1} = - \spb{\ell_1}{\ell_2} = m.
\end{equation}

\subsubsection{Fermion Wave Functions}
If the internal massive particle is a fermion, we must construct wave
functions which obey the Dirac equation.  Equation~(\ref{eq::ell1ell2})
which implies that
\begin{equation}
\begin{split}
\feynsl{\ell}\ket{\ell_1^+} = -m\,\ket{\ell_2^-}\qquad&
   \feynsl{\ell}\ket{\ell_2^-} = -m\,\ket{\ell_1^+}\,,\\
\feynsl{\ell}\ket{\ell_1^-} = \phantom{-}m\,\ket{\ell_2^+}\qquad&
   \feynsl{\ell}\ket{\ell_2^+} = \phantom{-}m\,\ket{\ell_1^-}\,,\\
\bra{\ell_1^+}\feynsl{\ell} = -m\,\bra{\ell_2^-}\qquad&
   \bra{\ell_2^-}\feynsl{\ell} = -m\,\bra{\ell_1^+}\,,\\
\bra{\ell_1^-}\feynsl{\ell} = \phantom{-}m\,\bra{\ell_2^+}\qquad&
   \bra{\ell_2^+}\feynsl{\ell} = \phantom{-}m\,\bra{\ell_1^-}\,,\\[5pt]
\end{split}
\end{equation}
which means that the Dirac spinors can be written as:
\begin{equation}
\begin{split}
\ket{u_{\uparrow}(\ell)} = \phantom{-}\ket{\ell_1^+} - \ket{\ell_2^-} \qquad&\
  \ket{u_{\downarrow}(\ell)} = \ket{\ell_2^+} + \ket{\ell_1^-}\,,\\
\ket{v_{\uparrow}(\ell)} = - \ket{\ell_2^+} + \ket{\ell_1^-} \qquad&\
  \ket{v_{\downarrow}(\ell)} = \ket{\ell_1^+} + \ket{\ell_2^-}\,,\\
\bra{\overline{u}_{\uparrow}(\ell)} =  \phantom{-}\bra{\ell_1^+} + \bra{\ell_2^-} \qquad&\
  \bra{\overline{u}_{\downarrow}(\ell)} = \bra{\ell_2^+} + \bra{\ell_1^-}\,,\\
\bra{\overline{v}_{\uparrow}(\ell)} = -\bra{\ell_2^+} + \bra{\ell_1^-} \qquad&\
  \bra{\overline{v}_{\downarrow}(\ell)} = \bra{\ell_1^+} + \bra{\ell_2^-}\,.\\
\end{split}
\end{equation}

Note that I label the spin states $\uparrow/\downarrow$, rather then
$\pm$.  This indicates that the decomposition of $\ell$ into $\ell_1$
and $\ell_2$ defined above does not yield wave functions that are
eigenstates of the helicity projection operator.  Since states with
complex momenta are necessarily internal states, I do not need
helicity projections, I only need to sum over the spin states.
One can easily verify that these spinors obey the Dirac Equation,
\begin{equation}
\begin{split}
\feynsl{\ell}\ket{u_{\uparrow}(\ell)} = \phantom{-}m\,\ket{u_{\uparrow}(\ell)} \qquad&\
  \feynsl{\ell}\ket{u_{\downarrow}(\ell)} = \phantom{-}m\,\ket{u_{\downarrow}(\ell)}\,,\\
\feynsl{\ell}\ket{v_{\uparrow}(\ell)} = -m\,\ket{v_{\uparrow}(\ell)} \qquad&\
  \feynsl{\ell}\ket{v_{\downarrow}(\ell)} = -m\,\ket{v_{\downarrow}(\ell)}\,,\\
\bra{\overline{u}_{\uparrow}(\ell)}\feynsl{\ell}
          = \phantom{-}m\,\bra{\overline{u}_{\uparrow}(\ell)} \qquad&\
  \bra{\overline{u}_{\downarrow}(\ell)}\feynsl{\ell}
          = \phantom{-}m\,\bra{\overline{u}_{\downarrow}(\ell)}\,,\\
\bra{\overline{v}_{\uparrow}(\ell)}\feynsl{\ell}
          = -m\,\bra{\overline{v}_{\uparrow}(\ell)} \qquad&\
  \bra{\overline{v}_{\downarrow}(\ell)}\feynsl{\ell}
          = -m\,\bra{\overline{v}_{\downarrow}(\ell)}\,,\\
\end{split}
\end{equation}
the standard normalization conditions,
\begin{equation}
\begin{split}
\braket{\overline{u}_{i}(\ell)}{u_{j}(\ell)} = \phantom{-}2\,m\,\delta_{ij}
   \quad& \braket{\overline{u}_{i}(\ell)}{v_{j}(\ell)} = 0\quad i,j\in\{\uparrow\,\downarrow\}\\
\braket{\overline{v}_{i}(\ell)}{v_{j}(\ell)} = {-}2\,m\,\delta_{ij}
   \quad& \braket{\overline{v}_{i}(\ell)}{u_{j}(\ell)} = 0\\
\end{split}
\end{equation}
and combine to form the standard projection operator,
\begin{equation}
\begin{split}
\label{eq:DiracProjector}
\sum_{i\in\{\uparrow\,\downarrow\}} \ket{u_{i}(\ell)}\bra{\overline{u}_{i}(\ell)} &= 
  \sum_{j\in\{+\,-\}}\Lx\ketbra{\ell_1^j}{\ell_1^j} + \ketbra{\ell_2^j}{\ell_2^j}\Rx\\
  + \ketbra{\ell_2^+}{\ell_1^-} +&\ketbra{\ell_1^-}{\ell_2^+}
  - \ketbra{\ell_1^+}{\ell_2^-} - \ketbra{\ell_2^-}{\ell_1^+}\\[5pt]
  &=\ \feynsl{\ell} + m\,,\\[10pt]
\sum_{i\in\{\uparrow\,\downarrow\}} \ket{v_{i}(\ell)}\bra{\overline{v}_{i}(\ell)} &=
  \sum_{j\in\{+\,-\}}\Lx\ketbra{\ell_1^j}{\ell_1^j} + \ketbra{\ell_2^j}{\ell_2^j}\Rx\\
  - \ketbra{\ell_2^+}{\ell_1^-} -&\ketbra{\ell_1^-}{\ell_2^+}
  + \ketbra{\ell_1^+}{\ell_2^-} + \ketbra{\ell_2^-}{\ell_1^+}\\[5pt]
  &=\ \feynsl{\ell} - m\,,\\
\end{split}
\end{equation}
where I can use the Schouten identities to show that
\begin{equation}
  \ketbra{\ell_2^+}{\ell_1^-} + \ketbra{\ell_1^-}{\ell_2^+} 
 - \ketbra{\ell_1^+}{\ell_2^-} - \ketbra{\ell_2^-}{\ell_1^+} = m\,.
\end{equation}
\begin{equation}
\begin{split}
-\spa{a}{\ell_1}\spa{\ell_2}{b} + \spa{a}{\ell_2}\spa{\ell_1}{b}
   &=\ -\spa{a}{b}\spa{\ell_2}{\ell_1} = m\,\spa{a}{b}\\
-\spb{a}{\ell_2}\spb{b}{\ell_1} + \spb{a}{\ell_1}\spb{\ell_2}{b}
   &=\ -\spb{a}{b}\spb{\ell_1}{\ell_2} = m\,\spb{a}{b}.
\end{split}
\end{equation}

In this way, I have defined orthogonal spin states for massive
fermions with complex momenta.  Although they are defined in terms of
momentum spinors of definite helicity, these spin states are not
eigenstates of the helicity projector.

\subsubsection{Massive Gauge Boson Wave Functions}
Massive gauge bosons have three physical spin states.  For outgoing
particles, the wave functions, or polarization vectors, can be written
as
\begin{equation}
\begin{split}
\varepsilon^{\uparrow\,\mu} &=\ \frac{\mtrxelm{\ell_1^+}{\gamma^\mu}{\ell_2^+}}
   {\sqrt{2}\spa{\ell_2}{\ell_1}} = -\frac{\mtrxelm{\ell_1^+}{\gamma^\mu}{\ell_2^+}}
   {\sqrt{2}\,m}\\
\varepsilon^{\downarrow\,\mu} &=\ \frac{\mtrxelm{\ell_1^-}{\gamma^\mu}{\ell_2^-}}
   {\sqrt{2}\spb{\ell_1}{\ell_2}} = -\frac{\mtrxelm{\ell_1^-}{\gamma^\mu}{\ell_2^-}}
   {\sqrt{2}\,m}\\
\varepsilon^{\bullet\,\mu} &=\ \frac{1}{m}\Lx\ell^{\mu}_1 - \ell^{\mu}_2\Rx
\end{split}
\end{equation}
As with the fermionic wave-functions, I do not use the standard symbols
of $\{+\,-\,0\}$ since these spin states are not eigenvectors of the
helicity projector.  They are, however, orthonormal and display all of
the usual properties expected of polarization vectors:
\begin{equation}
\begin{split}
\varepsilon^{\uparrow}\cdot\varepsilon^{\uparrow} =\ \phantom{-}0 \qquad &
\varepsilon^{\uparrow}\cdot\varepsilon^{\downarrow} =\ -1\qquad
\varepsilon^{\uparrow}\cdot\varepsilon^{\bullet} =\ \phantom{-}0\,,\\
\varepsilon^{\downarrow}\cdot\varepsilon^{\uparrow} =\ -1 \qquad &
\varepsilon^{\downarrow}\cdot\varepsilon^{\downarrow} =\ \phantom{-}0\qquad
\varepsilon^{\downarrow}\cdot\varepsilon^{\bullet} =\ \phantom{-}0\,,\\
\varepsilon^{\bullet}\cdot\varepsilon^{\uparrow} =\ \phantom{-}0 \qquad &
\varepsilon^{\bullet}\cdot\varepsilon^{\downarrow} =\ \phantom{-}0\qquad
\varepsilon^{\bullet}\cdot\varepsilon^{\bullet} =\ -1\,,\\
\varepsilon^{\uparrow\,\mu}\varepsilon^{\downarrow\,\nu}
    + \varepsilon^{\downarrow\,\mu}&\varepsilon^{\uparrow\,\nu}
    + \varepsilon^{\bullet\,\mu}\varepsilon^{\bullet\,\nu} = -
    g^{\mu\nu} + \frac{\ell^\mu\,\ell^\nu}{m^2}\\
\end{split}
\end{equation}


\subsection{Scalar Loop Integrals}
To establish my sign convention, I will define the scalar loop
integrals as follows,
\begin{equation}
\begin{split}
F_0&=\ (\ell^2 - m_0^2)\,,\\
F_1&=\ ((\ell+K_1)^2 - m_1^2)\,,\\
F_2&=\ ((\ell+K_2)^2 - m_2^2)\,,\\
F_3&=\ ((\ell+K_3)^2 - m_3^2)\,,\\
D_0&(m_0;K_1,m_1;K_2,m_2;K_3,m_3)\\
   &=\ -i(4\pi)^{D/2}\int\frac{d^D\ell}{(2\pi)^D}
    \frac{1}{F_0\,F_1\,F_2\,F_3}\,,\\
C_0&(m_0;K_1,m_1;K_2,m_2)=\ -i(4\pi)^{D/2}\int\frac{d^D\ell}{(2\pi)^D}
    \frac{1}{F_0\,F_1\,F_2}\,,\\
B_0&(m_0;K_1,m_1)=\ -i(4\pi)^{D/2}\int\frac{d^D\ell}{(2\pi)^D}
    \frac{1}{F_0\,F_1}\,,\\
A_0&(m_0)=\ -i(4\pi)^{D/2}\int\frac{d^D\ell}{(2\pi)^D}
    \frac{1}{F_0}\,.
\end{split}
\end{equation}
Note the sign convention on the momenta in the propagators.  If,
because of routing or some other convention, one of the momenta in the
denominator is negative, say $F_1 = \Lx(\ell-K_1)^2-m_1^2\Rx$, the
loop integral functions would be written as $B_0(m_0,-K_1,m_1)$,
$C_0(m_0,-K_1,m_1,K_2,m_2)$, etc.


\section{Methods}
\label{sec::methods}
In theories of with only massless particles propagating in loops, like
\qcd, it has been shown that any one-loop integral can be decomposed
into a sum of scalar box, triangle and bubble loop integral functions
that can be constructed from cuts using unitarity and a set of
rational terms~\cite{Bern:1994cg}
\begin{equation}
\begin{split}
A_{n,\,\qcd}^{1-loop} =\ \Rcal_n&\\
    + \frac{\mu^{2\ep}}{(4\pi)^{2-\ep}}&
     \Lx\sum_i b_i\,B_0^{(i)} + \sum_j c_j\,C_0^{(j)} + \sum_k d_k\,D_0^{(k)}\Rx\,,
\end{split}
\end{equation}
where $B_0^{(i)}$ represent the bubble integrals, $C_0^{(j)}$ the
triangle integrals and $D_0^{(k)}$ the boxes.  The fact that only box
functions are needed is because higher point functions can be written
as sums of the boxes formed by pinching vertices
together~\cite{Bern:1992em,Bern:1993kr}.  The rational terms,
$\Rcal_n$ cannot obtained from cuts, but can derived from on-shell
recursion relations~\cite{Britto:2004ap,Britto:2005fq,Bern:2005hs,
Bern:2005ji}

In theories with massive particles, this formula must be augmented to
include tadpole functions, $A_0^{(m)}$
\begin{equation}
\begin{split}
A_{n}^{1-loop} =\ \Rcal_n +
     \frac{\mu^{2\ep}}{(4\pi)^{2-\ep}}&
     \Lx\sum_m a_m\,A_0^{(m)} + \sum_i b_i\,B_0^{(i)}\right.\\
    & \left.+ \sum_j c_j\,C_0^{(j)} + \sum_k d_k\,D_0^{(k)}\Rx\,.
\end{split}
\end{equation}

In standard Passarino-Veltman reduction~\cite{Passarino:1978jh}, one
works in $D=4-2\ep$ dimensions and the coefficients of the loop
integral functions depend on the dimensional regulator $\ep$.
Rational terms develop when $\ep$-dependent pieces of the coefficients
multiply poles in $\ep$ from the loop integral function.  In the
methods described below, the cuts will be evaluated in $D=4$
dimensions, thereby missing those rational terms, which are assigned
to $\Rcal_n$.  $\Rcal_n$ will then be determined by loop-level
on-shell recursion relations as part of the unitarity bootstrap.

The basic idea of the unitarity method is that when one cuts a
one-loop amplitude, the terms on opposite sides of the cuts are
tree-level amplitudes.  These tree level amplitudes can be efficiently
computed using helicity methods.  However, helicity methods become
very messy when they are extended into $D=4-2\ep$ dimensions.  It is
desirable, therefore, to perform cuts in $D=4$ dimensions.  Doing so,
however, misses various rational terms, as described above, which must
be accounted for somehow.  In pure \qcd, a detailed knowledge of the
soft and collinear factorization properties of amplitudes is
sufficient to construct the missing rational terms.  In massive
theories, however, it is possible that there are rational terms that
cannot be constructed in this way~\cite{Bern:1995db}.  The development
of on-shell recursion
techniques~\cite{Britto:2004ap,Britto:2005fq,Bern:2005hs, Bern:2005ji}
to compute the rational terms allows the reliable use of four
dimensional unitarity cuts in massive theories as well.

In generalized unitarity, multiple cuts are made on the loop
amplitude, dividing it into multple tree amplitudes.  The method I
will describe below uses generalized unitarity to determine the
coefficients of loop integrals in terms of the tree-level amplitudes
that appear at the vertices of the cut diagrams.  Different approaches
to determining loop integral coefficients include algebraic
solution~\cite{Ossola:2006us,Ossola:2007ax}), using $D$-dimensional
cuts~\cite{Anastasiou:2006jv,Anastasiou:2006gt},and using standard
unitarity cuts via the holomorphic anomaly~\cite{Britto:2007tt}.


\section{Four Particle Cuts and the Extraction of Scalar Box Coefficients}
\label{sec::box}
\begin{figure}[ht]
\includegraphics[width=6.cm]{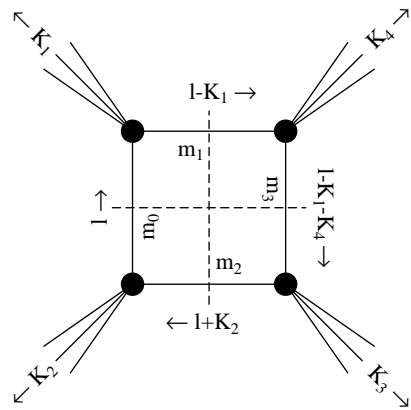}
\caption{a quadruple cut.}
\label{fig::quadcut}
\end{figure}

The use of quadruple cuts, shown in figure~\ref{fig::quadcut}, to extract the
box coefficient was first derived in \reference{Britto:2004nc}.
I re-derive the result here, using a more efficient momentum
parametrization and extending the procedure to explicitly permit
masses on the internal legs.  The basic procedure is to cut all four
internal legs, and put the cut legs on shell,
\begin{equation}
\frac{i}{(\ell+K_i)^2 - m_i^2}\longrightarrow
   (2\pi)\delta((\ell+K_i)^2 - m_i^2)
\end{equation}

\begin{figure}[ht]
\includegraphics[width=8.cm]{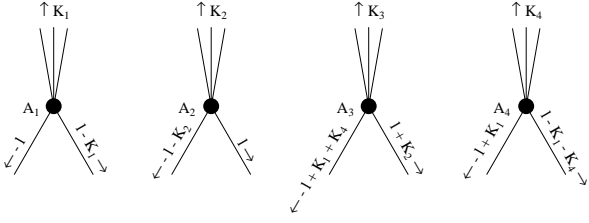}
\caption{Tree level amplitudes contributing to a particular box topology.}
\label{fig::quadamps}
\end{figure}

Imposing the cut conditions,
\begin{equation}
\begin{split}
\ell^2 - m_0^2 = 0\,,\hskip 37pt& \Lx\ell-K_1\Rx^2 - m_1^2 = 0\,,\\
\Lx\ell+K_2\Rx^2 - m_2^2 = 0\,,\hskip 10pt& \Lx\ell-K_1-K_4\Rx^2 -
m_3^2 = 0\,,
\end{split}
\end{equation}
singles out a unique box configuration. Other boxes satisfy a different
set of cuts and lower point diagrams cannot satisfy all four cuts.
The contribution of this box topology to the one-loop amplitude is
obtained by sewing together the four tree-level diagrams formed by
cutting the loop propagators (See figure~\ref{fig::quadamps}.)

In $D=4$ dimensions, the four delta-function constraints completely
determine the loop momentum $\ell^\mu$.  There are two solutions
and,in general, $\ell$ is complex valued.  The coefficient $d_m$,
corresponding to scalar box $D_0^{(m)}$, is given by averaging the
product of the four tree-level diagrams, evaluated with the solutions
to the loop momentum,
\begin{equation}
d_k =\ \frac{i}{2}\sum_{i=1}^2
  A^{(k)}_1(\ell_i)\,A^{(k)}_2(\ell_i)\,
  A^{(k)}_3(\ell_i)\,A^{(k)}_4(\ell_i)\,.
\end{equation}

\subsection{Parametrizing the Cut Loop Momentum}
It is convenient to parametrize $\ell$ in terms of the spinor
representations of two massless real momenta which can be constructed
from any two adjacent external momenta on the box.  (This
parametrization was previously used in
Refs.~\cite{delAguila:2004nf,Ossola:2006us,Ossola:2007ax}.)  I first
choose two adjacent external momenta, $K_1$ and $K_2$ and project them
onto one another to form two massless momenta, $K_1^\flat$ and
$K_2^\flat$.
\begin{equation}
\begin{split}
K_1 &=\ K_1^\flat + \frac{S_1}{\gamma_{12}} K_2^\flat\,,\qquad
K_2 =\ K_2^\flat + \frac{S_2}{\gamma_{12}} K_1^\flat\,,\\[5pt]
K_1^\flat &=\ \frac{K_1 - \frac{S_1}{\gamma_{12}}\,K_2}
            {1 - \frac{S_1\,S_2}{\gamma_{12}^2}}\,,\qquad
K_2^\flat =\ \frac{K_2 - \frac{S_2}{\gamma_{12}} K_1}
            {1 - \frac{S_1\,S_2}{\gamma_{12}^2}}\,,
\end{split}
\end{equation}
where
\begin{equation}
\begin{split}
S_1 &=\ K_1\cdot K_1\qquad S_2 =\ K_2\cdot K_2\,,\\
\gamma_{12} &= 2\,K_1^\flat\cdot K_2^\flat = K_1\cdot K_2 \pm \sqrt{\Delta(K_1,K_2)}\,,\\
\Delta(K_1,K_2) &=\ \Lx K_1\cdot K_2\Rx^2 - S_1\,S_2\,.
\end{split}
\end{equation}
Note that there are two solutions for $\gamma_{12}$ unless either
$S_1=0$ or $S_2=0$.

Assuming that the Gram determinant, $\Delta(K_1,K_2)$, does not
vanish, I can now form four massless vectors to use as a basis for
solving for the loop momentum $\ell$.
\begin{equation}
\begin{split}
a_1^\mu &= K_1^{\flat\,\mu}\,,\qquad\quad\qquad\qquad a_2^\mu = K_2^{\flat\,\mu}\,,\\
a_3^\mu &= \mtrxelm{K_1^{\flat\,-}}{\gamma^\mu}{K_2^{\flat\,-}}\,,\qquad
     a_4^\mu =
     \mtrxelm{K_2^{\flat\,-}}{\gamma^\mu}{K_1^{\flat\,-}}\,,\\
\ell^\mu &= \alpha_1\,a_1^\mu + \alpha_2\,a_2^\mu + \alpha_3\,a_3^\mu
+ \alpha_4\,a_4^\mu\,.
\end{split}
\end{equation}
This basis has the property that $a_3\cdot a_4 = 4\,a_1\cdot a_2 =
2\gamma_{12}$, but all other inner products of the $a_i$ among themselves
vanish.  The solutions for $\ell$ are:
\begin{equation}
\begin{split}
\alpha_1 &=\ \frac{S_2\Lx\gamma_{12}-S_1\Rx + \Lx\gamma_{12}-S_2\Rx\,m_0^2 -
    \gamma_{12}\,m_2^2 + S_2\,m_1^2}{\gamma_{12}^2 - S_1\,S_2}\,,\\
\alpha_2 &=\ \frac{S_1\Lx\gamma_{12}-S_2\Rx + \Lx\gamma_{12}-S_1\Rx\,m_0^2 -
    \gamma_{12}\,m_1^2 + S_1\,m_2^2}{\gamma_{12}^2 - S_1\,S_2}\,,\\
\beta_3 &=\  2\Lx\alpha_1-1\Rx K_1^\flat\cdot K_4
    + 2\Lx\alpha_2-\frac{S_1}{\gamma_{12}}\Rx K_2^\flat\cdot K_4\\
    &\hskip 20pt - S_4
    + m_3^2 - m_1^2\,,\\
\beta_4 &=\ \alpha_1\,\alpha_2 - \frac{m_0^2}{\gamma_{12}}\,,\\
\alpha_3 &=\ -\frac{\beta_3\pm\sqrt{\beta_3^2
     - 2\beta_4\Tr{\feynsl{K}_1^\flat\,\feynsl{K}_4
         \,\feynsl{K}_2^\flat\,\feynsl{K}_4}}}
        {2\mtrxelm{K_1^{\flat\,-}}{\feynsl{K}_4}{K_2^{\flat\,-}}}\,,\\
\alpha_4 &=\ \frac{\beta_4}{4\,\alpha_3}\,.
\label{eq::quadloopsoln}
\end{split}
\end{equation}
There are two solutions for $\alpha_3$ and it might appear that,
combined with the two solutions for $\gamma_{12}$, there are four
solutions for $\ell$.  However, it works out that
\begin{equation}
\begin{split}
\ell^\mu(\gamma^+_{12},\alpha^+_3) &=\ \ell^\mu(\gamma^-_{12},\alpha^-_3)\,,\\
\ell^\mu(\gamma^+_{12},\alpha^-_3) &=\ \ell^\mu(\gamma^-_{12},\alpha^+_3)\,,
\end{split}
\end{equation}
so that there are only two solutions for $\ell^\mu$.

This parametrization of $\ell^\mu$ looks rather different than that in
\reference{Britto:2004nc}, which used as a basis the three independent
external momenta, $K_1^\mu$, $K_2^\mu$ and $K_4^\mu$, and the
antisymmetric combination of those three, $P^\mu =
\varepsilon^{\mu\nu\rho\lambda}\,K_{1\,\nu}\,K_{2\,\rho}\,K_{4\,\lambda}$.
Numerically, of course, the solutions are identical, but that given
here has a number of features to recommend it.  First, it is quite
compact and easy to compute, even allowing for arbitrary values for
the internal and external masses.  Second, it is better behaved in
Gram-singular configurations.  As with all integral reduction
techniques, we find spurious singularities in the form of inverse
powers of the Gram determinant arising from tensor reduction.  The
Gram determinants, $\Delta(K_1,K_2,K_4)$, are found in the on-shell
solution to the loop momentum $\ell$.  Numerical analysis shows that
$\ell$ scales like $1/\sqrt{\Delta(K_1,K_2,K_4)}$ near the Gram
singularity.  This property is explicit in \eqn{eq::quadloopsoln},
where the denominator of $\alpha_3$,
$\mtrxelm{K_1^{\flat\,-}}{\feynsl{K}_4}{K_2^{\flat\,-}}$ is the
(complex) square root of $\Delta(K_1,K_2,K_4)$.  In the solution of
\reference{Britto:2004nc}, the components of $\ell$ have coefficients
that scale like $1/P^2$, where $P^2 = \Delta(K_1,K_2,K_4)$.  Since
$\ell$ actually scales like $1/\sqrt{\Delta(K_1,K_2,K_4)}$, there are
cancellations hidden in the parametrization.  This makes the problem
of locating and canceling these spurious singularities much more
difficult.


\section{Three Particle Cuts and the Extraction of Scalar Triangle Coefficients}
\label{sec::triangle}

The procedure for extracting the box coefficients is so remarkably
simple that one is inspired to try the approach for extracting the
coefficients of lower point integrals.  The use of triple cuts in one
loop amplitudes to extract the triangle coefficient was first
discussed by Mastrolia~\cite{Mastrolia:2006ki}.  A complication arises
when one imposes a triple cut on a one-loop amplitude.  One does pick
out a single triangle topology, but one also gets contributions from
box topologies formed by splitting one of the triangle's vertices to
open up a fourth propagator.  Since the box contributions can be
determined by the prescription above, one needs a way of separating
the pure triangle contributions from the already known box terms.
Forde~\cite{Forde:2007mi} has recently described an elegant way of
doing so.  The discussion below follows that of Forde, but allows for
massive internal legs on the triangles.

\begin{figure}[ht]
\includegraphics[width=6.cm]{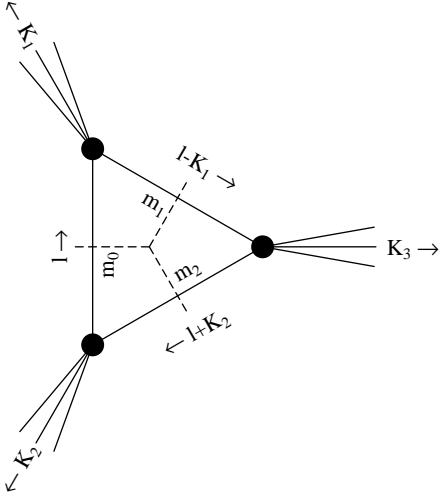}
\caption{a triple cut.}
\label{fig::triplecut}
\end{figure}

I consider a triple-cut triangle, as shown in
figure~\ref{fig::triplecut}.  The three delta function constraints
imposed by the cuts,
\begin{equation}
\delta(\ell^2 - m_0^2)\,,\ \delta(\Lx\ell-K_1\Rx^2 - m_1^2)\,,\ 
\delta(\Lx\ell+K_2\Rx^2 - m_2^2)\,,
\end{equation}
are not sufficient to completely fix the loop
momentum, $\ell$, which must therefore have an unconstrained degree of
freedom.  Using the same basis involving $K_1^\flat$ and $K_2^\flat$
as for the box, I can parametrize the loop momentum as
\begin{equation}
\begin{split}
\ell^\mu &=\ \alpha_1\,a_1^\mu + \alpha_2\,a_2^\mu +
   \frac{t}{2}a_3^\mu + \frac{\alpha_4}{2\,t}a_4^\mu\,,\\
\alpha_1 &=\ \frac{S_2\Lx\gamma_{12}-S_1\Rx + \Lx\gamma_{12}-S_2\Rx\,m_0^2 -
    \gamma_{12}\,m_2^2 + S_2\,m_2^2}{\gamma_{12}^2 - S_1\,S_2}\,,\\
\alpha_2 &=\ \frac{S_1\Lx\gamma_{12}-S_2\Rx + \Lx\gamma_{12}-S_1\Rx\,m_0^2 -
    \gamma_{12}\,m_1^2 + S_1\,m_2^2}{\gamma_{12}^2 - S_1\,S_2}\,,\\
\alpha_4 &=\ \alpha_1\,\alpha_2 - \frac{m_0^2}{\gamma_{12}}\,.
\end{split}
\end{equation}

The integral depicted in figure~\ref{fig::triplecut} is given by
\begin{equation}
\begin{split}
c_j&\,C_0(m_0,-K_1,m_1,K_2,m_2)\\
 =\ &i\,\int \frac{d^4\ell}{(2\pi)^4}\frac{A^{(j)}_1(K_1;\ell)\,A^{(j)}_2(K_2;\ell)\,
    A^{(j)}_3(K_3;\ell)}{(\ell^2 - m_0^2)(\Lx\ell-K_1\Rx^2 - m_1^2)
    (\Lx\ell+K_2\Rx^2 - m_2^2)}\\
\to&i\,(-2\pi\,i)^3\int \frac{d^4\ell}{(2\pi)^4}A^{(j)}_1(K_1;\ell)\,
    A^{(j)}_2(K_2;\ell)\,A^{(j)}_3(K_3;\ell)\,\times\\
&\hskip 20pt\delta(\ell^2 - m_0^2)\,
    \delta(\Lx\ell-K_1\Rx^2 - m_1^2)\,\delta(\Lx\ell+K_2\Rx^2 -
    m_2^2)\\
&=\ i\,(-2\pi\,i)^3\int\frac{dt}{(2\pi)^4}\,J_t\,A^{(j)}_1(t)\,A^{(j)}_2(t)\,A^{(j)}_3(t)\,,
\end{split}
\end{equation}
where $J_t$ is the Jacobian of the transformation from the
delta-function constrained integral over $d^4\ell$ to the integral
over the remaining free parameter $t$.  Treating $t$ as a complex
variable and partial fractioning off terms with poles at finite $t$,
this last integral can be rewritten as
\begin{equation}
\begin{split}
c_j =i\,(-2\pi\,i)^3\int\frac{dt}{(2\pi)^4}&\,J_t\,\Lx
  \Inf{t}{A^{(j)}_1\,A^{(j)}_2\,A^{(j)}_3}(t)\right.\\ 
 +  &\left.\sum_{\{k\}}\Res{t=t_k}{A^{(j)}_1\,A^{(j)}_2\,A^{(j)}_3}{t-t_k}\Rx\,,
\end{split}
\end{equation}
which represents a sum over the residues of all poles $\{k\}$ at
finite $t_k$ and a contribution at infinity.  The $\Inf{t}{}$ term is
defined so that
\begin{equation}
\lim_{t\to\infty}\Lx\Inf{t}{A_1\,A_2\,A_3}(t) -
   A_1(t)\,A_2(t)\,A_3(t)\Rx = 0\,.
\end{equation}
The $\Inf{t}{}$ term will be some polynomial in $t$,
\begin{equation}
\Inf{t}{A_1\,A_2\,A_3}(t) = \sum_{n=0}^m e_n\,t^n\,,
\end{equation}
where $m$ is set by the maximum tensor rank allowed.  For
renormalizable theories, the maximum tensor rank for the triangle is
three.

Now, recall from the discussion of the box coefficient that the fourth
delta function constraint fixed the value of $t$ at some complex value
$t_0$.  Thus, the sum over residues at finite $t$ simply correspond to
the contributions to the triple-cut from the various box
configurations that satisfy those three cuts.  Moreover, the terms
only give contributions to the scalar box coefficients.  The triangle
contributions of the triple-cut come exclusively from the terms at
infinity.  That is,
\begin{equation}
c_j =\ i\,(-2\pi\,i)^3\int\frac{dt}{(2\pi)^4}\,J_t\,
   \Inf{t}{A^{(j)}_1\,A^{(j)}_2\,A^{(j)}_3}(t)\,.
\end{equation}

So now, I need to construct a table for integrals of the form
\begin{equation}
i\,(-2\pi\,i)^3\int\frac{dt}{(2\pi)^4}\,J_t\,t^n\,.
\end{equation}
This is easily done by considering tensor triangle integrals of the
form
\begin{equation}
\begin{split}
&\int\frac{d^4\ell}{(2\pi)^4}\,\frac{\Lx\ell\cdot a_4\Rx^n}
   {(\ell^2 - m_0^2)(\Lx\ell-K_1\Rx^2 - m_1^2)
    (\Lx\ell+K_2\Rx^2 - m_2^2)}\\
\to&(-2\pi\,i)^3\,\gamma_{12}^n\int\frac{dt}{(2\pi)^4}\,J_t\,t^n
\end{split}
\end{equation}
For the scalar integral, $n=0$, it is clear that
\begin{equation}
i\,(-2\pi\,i)^3\int\frac{dt}{(2\pi)^4}\,J_t = 1.
\end{equation}
For tensor triangles however, with $n>0$, the integrals vanish.  By
Lorentz invariance, the components of the tensor triangles must
involve only products of $K_1$, $K_2$ or the metric tenor
$g^{\mu\nu}$.  But $K_1$ and $K_2$ decompose into $K_1^\flat$ and
$K_2^\flat$, which annihilate $a_4^\mu$ via the Dirac equation, and
metric tensor terms vanish because $a_4\cdot a_4 = 0$.  Therefore,
\begin{equation}
i\,(-2\pi\,i)^3\int\frac{dt}{(2\pi)^4}\,J_t\,t^{n>0} = 0\,,
\end{equation}
and
\begin{equation}
c_j =\ -\Inf{t}{A^{(j)}_1\,A^{(j)}_2\,A^{(j)}_3}(t)|_{t=0}\,.
\end{equation}


\section{Two Particle Cuts and the Extraction of Scalar Bubble Coefficients}
\label{sec::bubble}

To extract the coefficients of bubble integrals, I proceed as before
and impose the cuts that define the bubble topology
\begin{equation}
\delta(\ell^2 - m_0^2)\,,\qquad \delta((\ell-K_1)^2 - m_1^2)\,.
\end{equation}
Only one bubble configuration will satisfy these cuts, but multiple
triangle and box configurations will do so.  Since the boxes and
triangles can be extracted by the methods above, the task here is to
isolate the pure bubble contributions.

\begin{figure}[ht]
\includegraphics[width=6.cm]{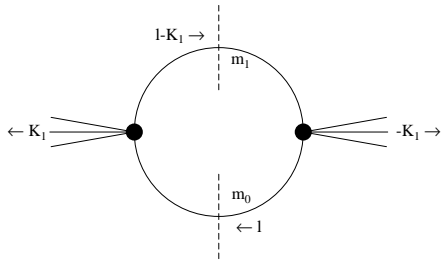}
\caption{a double cut.}
\label{fig::doublecut}
\end{figure}

Since I only have one external momentum, $K_1$, in a bubble
configuration, I will choose an arbitrary massless momentum $\chi^\mu$
to define my parametrization.  My massless projection of $K_1$ and
the set of basis vectors for the loop momentum is then defined in
terms of $\chi^\mu$:
\begin{equation}
\begin{split}
K_1 &=\ K_1^\flat + \frac{S_1}{\gamma_{1\chi}}\chi\,,\hskip 35pt
   \gamma_{1\chi} = 2\,K_1\cdot\chi = 2\,K_1^\flat\cdot\chi\,,\\
a_1^\mu &= K_1^{\flat\,\mu}\,,\hskip 65pt a_2^\mu = \chi^\mu\,,\\
a_3^\mu &= \mtrxelm{K_1^{\flat\,-}}{\gamma^\mu}{\chi^-}\,,\qquad
     a_4^\mu =
     \mtrxelm{\chi^-}{\gamma^\mu}{K_1^{\flat\,-}}\,,\\
\ell^\mu &= y\,a_1^\mu + \alpha_2\,a_2^\mu + \frac{t}{2}\,a_3^\mu
+ \alpha_4\,a_4^\mu\,,
\end{split}
\end{equation}
where both $y$ and $t$ are free parameters.  The other coefficients,
$\alpha_2$ and $\alpha_4$ are fixed by the cut conditions and are
found to be
\begin{equation}
\alpha_{\,2} =\ \frac{S_1 + m_0^2 - m_1^2}{\gamma_{1\chi}}
     - \frac{S_1}{\gamma_{1\chi}}y\,,\quad
  \alpha_4 = \frac{1}{2\,t}\Lx\alpha_{\,2}\,y - \frac{m_0^2}{\gamma_{1\chi}}\Rx\,.
\end{equation}

The integral depicted in figure~\ref{fig::doublecut} is
\begin{equation}
\begin{split}
b_i&\,B_0(m_0,-K_1,m_1)\\
 =\ &i\int \frac{d^4\ell}{(2\pi)^4}\frac{A^{(i)}_1(K_1;\ell)\,A^{(i)}_2(K_2;\ell)}
   {(\ell^2 - m_0^2)\Lx\Lx\ell-K_1\Rx^2 - m_1^2\Rx}\\
\to\ &i\,(-2\pi\,i)^2\int \frac{d^4\ell}{(2\pi)^4}A^{(i)}_1(K_1;\ell)\,
    A^{(i)}_2(K_2;\ell)\,\times\\
&\hskip 20pt\delta\Lx\ell^2 - m_0^2\Rx\,
    \delta\Lx\Lx\ell-K_1\Rx^2 - m_1^2\Rx\\
=\ &i\,(-2\pi\,i)^2\int\frac{dt\,dy}{(2\pi)^4}\,J_{t,y}\,A^{(i)}_1(t,y)\,A^{(i)}_2(t,y)\,,
\end{split}
\end{equation}
where $J_{t,y}$ is the Jacobian of the transformation.  This time
treating both $y$ and $t$ as complex variables and applying partial
fractioning, this last integral can be rewritten as:
\begin{equation}
\begin{split}
\label{eq::twocut}
b_i = i\,(-2\pi\,i)^2&\int\frac{dt\,dy}{(2\pi)^4}\,J_{t,y}\,A_1^{(i)}(t,y)\,A^{(i)}_2(t,y)
  \\
\to\ i\,(-2\pi\,i)^2\int&\frac{dt\,dy}{(2\pi)^4}\,J_{t,y}\Lx\Inf{y}{\Inf{t}{A^{(i)}_1\,A^{(i)}_2}}(t,y)\right.\\
   &+\Inf{y}{\sum_{\{k\}}\Res{t=t_k}{A^{(i)}_1\,A^{(i)}_2}{t-t_k}}(y)\\
   &+\sum_{\{j\}}\Res{y=y_j}{\Inf{t}{A^{(i)}_1\,A^{(i)}_2}(t)}{y-y_j}\\
   &+\sum_{\{j\}}\Res{y=y_j}{\sum_{\{k\}}\Res{t=t_k}{A^{(i)}_1\,A^{(i)}_2}{t-t_k}}{y-y_j}\\
\end{split}
\end{equation}
The double $\Inf{}{}$ terms are the pure bubble contributions to the
double-cut, the single residue terms are triangle contributions and
the double residue terms are box contributions.  As before, the box
contributions only give the scalar box coefficients and are therefore
not of interest to extracting bubble coefficients.  The triangle terms
are not so easily dismissed.  Unlike in the previous section, the
parametrization of the loop momentum does not annihilate all tensor
triangle contributions.  Since tensor triangles can be decomposed into
scalar triangles, bubbles and tadpoles, there is a contribution to the
bubble coefficient from the single residue terms.

\subsection{Pure Bubble Contributions to the Bubble Coefficient}
To extract the pure bubble contribution to the coefficient, I must
build an integral table for powers of both $t$ and $y$.  If I consider
tensor bubbles of the form
\begin{equation}
\begin{split}
&\int\frac{d^4\ell}{(2\pi)^4}\frac{\Lx\ell\cdot a_4\Rx^n}
   {(\ell^2 - m_0^2)(\Lx\ell-K_1\Rx^2 - m_1^2)}\\
\to&(-2\pi\,i)^2\,\gamma_{1\chi}^n\int\frac{dt\,dy}{(2\pi)^4}J_{t,y}\ t^n
\end{split}
\end{equation}
For $n=0$, this is just the scalar bubble and the result is of course
equal to unity.  For $n>0$, the integrals vanish because, by Lorentz
invariance, tensor bubbles have components made up of $K_1^{\mu_i}$ and
the metric tensor $g^{\mu_i\mu_j}$, both of which annihilate products of
$a_4^{\mu_1}\dots a_4^{\mu_n}$.  Thus,
\begin{equation}
\begin{split}
\int\frac{dt\,dy}{(2\pi)^4}J_{t,y}& =\ 1\,,\\
\int\frac{dt\,dy}{(2\pi)^4}J_{t,y}&\ t^{n>0} =\ 0\,.
\end{split}
\end{equation}

The integral table for values of $y$ is somewhat more complicated.
First, I form the auxiliary vector
\begin{equation}
\begin{split}
\omega &= K_1^\flat - \frac{S_2}{\gamma_{1\chi}}\chi\,,\qquad
K_1\cdot\omega = 0\,,\qquad \omega^2 = - S_1\,,\\
\ell\cdot\omega &= \frac{1}{2}\Lx S_1+m_0^2-m_1^2\Rx - S_1\,y
\end{split}
\end{equation}
Using $\omega$, I find,
\begin{equation}
\begin{split}
\int\frac{d^4\ell}{(2\pi)^4}&\frac{\Lx\ell\cdot\omega\Rx^{2k-1}}
   {\Lx\ell^2-m_0^2\Rx\,\Lx(\ell-K_1)^2 - m_1^2\Rx} = 0\\
   \Rightarrow \int& dt\,dy\,J_{t,y}\Lx\frac{S_1 + m_0^2 -
   m_1^2}{2} - S_1\,y\Rx^{2k-1} = 0\,,\\
\int\frac{d^4\ell}{(2\pi)^4}&\frac{\Lx\ell\cdot\omega\Rx^{2k}}
   {\Lx\ell^2-m_0^2\Rx\,\Lx(\ell-K_1)^2 - m_1^2\Rx} =
     \frac{(2k)!}{2^k\,k!}(-S_1)^{k}\,B^{\{0^{2k}\}}\\
   \Rightarrow \int& dt\,dy\,J_{t,y}\Lx\frac{S_1 + m_0^2 -
   m_1^2}{2\,S_1} - y\Rx^{2k} \\
   &\hskip 50pt= \Lx\frac{-1}{S_1}\Rx^{k}\frac{(2k)!}{2^k\,k!}
     \widetilde{B}^{\{0^{2k}\}}\,,
\end{split}
\end{equation}
where $B^{\{0^{2k}\}}$ is the Passarino-Veltman reduction term
containing only products of the metric tensor (no factors of
$K_1^\mu$), and $\widetilde{B}^{\{0^{2k}\}}$ is the corresponding
coefficient of the scalar bubble in that term.  One can avoid actually
constructing the Passarino-Veltman reduction by using another set of
identities,
\begin{equation}
\begin{split}
\int\frac{d^4\ell}{(2\pi)^4}&\frac{\Lx\ell\cdot a_3\Rx^{k}\Lx\ell\cdot a_4\Rx^{k}}
   {\Lx\ell^2-m_0^2\Rx\,\Lx(\ell-K_1)^2 - m_1^2\Rx} = (-1)^k\,2^k\,k!\,\gamma_{1\chi}^{k}
      \,B^{\{0^{2k}\}}\\
   \Rightarrow& \int dt\,dy\,J_{t,y}\,\Lx -\frac{m_0^2}{S_1}
        + y\frac{S_1 + m_0^2 - m_1^2}{S_1} - y^2\Rx^{k} \\
   &\hskip 50pt = \Lx\frac{-1}{S_1}\Rx^k\,2^k\,k!
      \,\widetilde{B}^{\{0^{2k}\}}\,,
\end{split}
\end{equation}
Eliminating the Passarino-Veltman coefficient, I find
\begin{equation}
\begin{split}
\int\frac{dt\,dy}{(2\pi)^4}J_{t,y}& =\ 1\,,\\
\int\frac{dt\,dy}{(2\pi)^4}J_{t,y}&\Lx\frac{S_1 + m_0^2 -
   m_1^2}{2} - S_1\,y\Rx^{2k-1} = 0\,,\\
\int\frac{dt\,dy}{(2\pi)^4}J_{t,y}&\LB\frac{2^k\,k!}{(2k)!}\Lx\frac{S_1 + m_0^2 -
   m_1^2}{2\,S_1} - y\Rx^{2k}\right.\\
  -&\left.\frac{1}{2^k\,k!}\Lx -\frac{m_0^2}{S_1}
        + y\frac{S_1 + m_0^2 - m_1^2}{S_1} - y^2\Rx^{k}\RB=0\,.
\end{split}
\end{equation}
From these expressions, I can build an integral table for $y$:
\begin{equation}
\begin{split}
 B(n)&=\int\frac{dt\,dy}{(2\pi)^4}J_{t,y}\ y^n\\
 &= \frac{(-1)^n}{n+1}\sum_{i=0}^{\lfloor n/2\rfloor}
   \binom{n-i}{i}\Lx\frac{-m_0^2}{S_1}\Rx^i\Lx\frac{S_1+m_0^2-m_1^2}{S_1}
   \Rx^{n-2i}\,.
\end{split}
\end{equation}

\subsection{Triangle Contributions to the Bubble Coefficient}
It is also possible for there to be triangle contributions to the
bubble coefficient.  These will come from the single residue terms in
\eqn{eq::twocut}.  The residue terms correspond to an additional
propagator going on-shell.  Therefore, these terms can be obtained by
applying an additional constraint to the double cut of the form,
\begin{equation}
\delta\Lx(\ell+K_2)^2 - m_2^2\Rx\,.
\end{equation}
Without any loss of generality, $y$ can be eliminated to satisfy the
constraint, leaving $t$ as the sole free parameter.  The equation in
$y$ is
\begin{equation}
\begin{split}
0=y^2\,\frac{S_1}{t\,\gamma_{1\chi}}&\mtrxelm{\chi^-}{\feynsl{K}_2}{K_1^{\flat\,-}}\\
   -&2\,y\LB\,K_1^\flat\cdot K_2 -
    \frac{S_1}{\gamma_{1\chi}}\chi\cdot K_2\right.\\
    &\left.+\frac{S_1+m_0^2-m_1^2}{2\,t\,\gamma_{1\chi}}\mtrxelm{\chi^-}
   {\feynsl{K}_2}{K_1^{\flat\,-}}\RB\\
   -&2\frac{S_1+m_0^2-m_1^2}{\gamma_{1\chi}}\chi\cdot K_2 + S_2 +
   m_0^2-m_2^2\\
   +&t\mtrxelm{K_1^{\flat\,-}}{\feynsl{K}_2}{\chi^-}
    -\frac{m_0^2}{t\,\gamma_{1\chi}}\mtrxelm{\chi^-}{\feynsl{K}_2}{K_1^{\flat\,-}}
\end{split}
\end{equation}
There are two solutions, $y=y_\pm$, which must be averaged over.
Because the parametrization of the loop parameter is based upon the
vectors $K_1$ and $\chi$, instead of $K_1$ and $K_2$, the tensor
components of the triangle, that is, non-zero powers of $t$, will not
identically vanish.

The integral table for $t$ can be worked out by considering tensor
integrals of the form
\begin{equation}
\begin{split}
&\int\frac{d^4\ell}{(2\pi)^4}\frac{\Lx\ell\cdot a_4\Rx^n}
   {(\ell^2 - m_0^2)\Lx(\ell-K_1)^2 - m_1^2\Rx
   \Lx(\ell+K_2)^2 - m_2^2\Rx}\\
\to&(-2\pi\,i)^3\,\gamma_{1\chi}^n\int\frac{dt\,dy}{(2\pi)^4}J^\prime_{t}\ t^n\,.
\end{split}
\end{equation}
Because $a_4$ is a null vector and is orthogonal to $K_1$, if I
perform Passarino-Veltman reduction, I find that only the $\Lx
K_2^\mu\Rx^n$ component contributes to this integral.  I then pick out
the coefficient of the particular bubble I am looking at,
$B_0(m_0,-K_1,m_1)$, from that Passarino-Veltman term.  I find that
\begin{equation}
\begin{split}
T(n) &= \int\frac{dt\,dy}{(2\pi)^4}J^\prime_{t}\
  t^n\\
  &= -\Lx\frac{S_1}{2\,\gamma_{1\chi}}\Rx^n
  \frac{\mtrxelm{\chi^-}{\feynsl{K}_2}{K_1^{\flat\,-}}^n\,\Lx K_1\cdot
  K_2\Rx^{n-1}}{\Delta^n(K_1,K_2)}{\cal C}_n\,,\\
\end{split}
\end{equation}
where $\Delta(K_1,K_2)$ is the triangle Gram determinant and
\begin{equation}
\begin{split}
\label{eq::bubxtri}
{\cal C}_0 &= 0\,,\\
{\cal C}_1 &= 1\,,\\
{\cal C}_2 &= \frac{3}{2}\Lx\frac{S_1+m_0^2-m_1^2}{S_1}
    - \frac{S_2+m_0^2-m_2^2}{K_1\cdot K_2}\Rx\,,\\
{\cal C}_3 &= \frac{5}{2}\Lx\frac{S_1+m_0^2-m_1^2}{S_1}
    - \frac{S_2+m_0^2-m_2^2}{K_1\cdot K_2}\Rx^2\\
    &\qquad-\frac{2}{3}\frac{\Delta(K_1,K_2)}{\Lx K_1\cdot K_2\Rx^2}
     \LB\Lx\frac{S_1+m_0^2-m_1^2}{S_1}\Rx^2 - 4\frac{m_0^2}{S_1}\RB\,.
\end{split}
\end{equation}

\subsection{The Bubble Coefficient}
The complete bubble coefficient is found by combining the pure bubble
and triangle contributions.
\begin{equation}
\begin{split}
b_i =\ &
-i\,\left.\Inf{y}{\Inf{t}{A^{(i)}_1\,A^{(i)}_2}}(t,y)\right|_{t\to0\,,y^n\to
  B(n)}\\
 &-\frac{1}{2}\sum_{\rm triangles}\sum_{y=y_\pm}\left.\Inf{t}{
   \tilde{A}^{(i)}_1\,\tilde{A}^{(i)}_2\,\tilde{A}^{(i)}_3}(t)
   \right|_{t^n\to T(n)}\,,
\end{split}
\end{equation}
where the $\tilde{A}^{(i)}_n$ are the amplitudes formed by cutting one
more propagator in either $A^{(i)}_1$ or $A^{(i)}_2$.


\section{Single Particle Cuts and the Extraction of Scalar Tadpole Coefficients}
\label{sec::tadpole}

The tadpole coefficients can be extracted by an extension of the same
procedure.  The only constraint that the loop momentum must satisfy is
that $\ell^2 - m_0^2 = 0$.  Satisfying this constraint leaves three
free parameters in $\ell$,
\begin{equation}
\begin{split}
\label{eq::tadcut}
\ell^\mu &= y\,\chi^\mu + w\,\psi^\mu + \frac{t}{2}\,\mtrxelm{\chi^-}{\gamma^\mu}{\psi^-}
+ \alpha_4\,\mtrxelm{\psi^-}{\gamma^\mu}{\chi^-}\,,\\
\alpha_4 &= \frac{1}{2\,t}\Lx w\,y - \frac{m_0^2}{\gamma_{\chi\psi}}\Rx\,,
    \hskip 55pt \gamma_{\chi\psi} = 2\,\chi\cdot\psi\,,\\
\end{split}
\end{equation}
where $\chi^\mu$ and $\psi^\mu$ are arbitrary light-like momenta,
since, in a tadpole configuration, there are no external momenta on
which to base the parametrization.  

\begin{figure}[ht]
\includegraphics[width=5.cm]{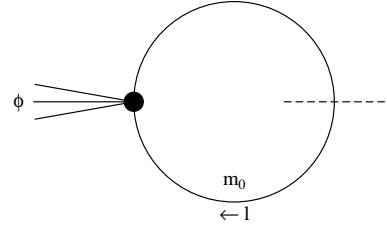}
\caption{a single cut.}
\label{fig::singlecut}
\end{figure}

The integral depicted in figure~\ref{fig::singlecut} is
\begin{equation}
\begin{split}
a_m\,A_0(m_0)&=\ i\,\int \frac{d^4\ell}{(2\pi)^4}\frac{A^{(m)}(\emptyset;\ell)}
   {(\ell^2 - m_0^2)}\\
&\to\ i\,(-2\pi\,i)\int \frac{d^4\ell}{(2\pi)^4}A^{(m)}(\emptyset;\ell)
    \delta\Lx\ell^2 - m_0^2\Rx\\
&=\ i\,(-2\pi\,i)\int\frac{dt\,dy\,dw}{(2\pi)^4}\,J_{t,y,w}\,A^{(m)}(t,y,w)\,,
\end{split}
\end{equation}
where $A(\emptyset;\ell)$ indicates that there is no external momentum
flowing in to the tadpole.  I again treat the free parameters as
complex variables and apply partial fractioning to obtain,
\begin{equation}
\begin{split}
\label{eq::onecut}
(-2\pi\,i)\int&\frac{dt\,dy\,dw}{(2\pi)^4}\,J_{t,y,w}\,A(t,y,w)
  =\\
(-2\pi\,i)\int&\frac{dt\,dy\,dw}{(2\pi)^4}\,J_{t,y,w}\times\\
   &\qquad\Lx\Inf{w} {\Inf{y}{\Inf{t}{A}}}(t,y,w) + \dots\Rx\,.
\end{split}
\end{equation}
The triple $\Inf{}{}$ term will be the pure tadpole contribution to
the tadpole, the double $\Inf{}{}$ - single $\LB{\rm Res}\RB$ terms will be
the bubble contribution to the tadpole and the single $\Inf{}{}$ -
double $\LB{\rm Res}\RB$ terms will be the triangle contribution to the
tadpole.  As before, the pure residue terms correspond to scalar box
contributions and need not be considered.

\subsection{Pure Tadpole Contributions to the Tadpole}
To extract the pure tadpole contribution to the coefficient, I must
build an integral table for powers of $w$, $y,$ and $t$.  This task is
greatly simplified by the fact that tensor tadpole integrals vanish
when the rank of the tensor is odd.  Since there are no external
momenta to project upon, the only non-vanishing tensors are
proportional to products of the metric tensor.  Further
simplifications follow from the fact that the basis vectors are
light-like and the only non-vanishing contractions are
\begin{equation}
\mtrxelm{\chi^-}{\gamma^\mu}{\psi^-}\,\mtrxelm{\psi^-}{\gamma_\mu}{\chi^-}
   = -4 \chi\cdot\psi\,.
\end{equation}
These facts lead to the conclusion that the only non-vanishing entries
in the integral table are of the form
\begin{equation}
\int\frac{dt\,dy\,dw}{(2\pi)^4}\,J_{t,y,w}\ y^n\,w^n\ne 0\,,
\end{equation}
where $n\ge0$.  All integrals involving non-zero powers of $t$ vanish
identically.
\begin{equation}
\begin{split}
\int\frac{d^4\ell}{(2\pi)^4}&\frac{(\ell\cdot\chi)^k(\ell\cdot\psi)^k}
   {\ell^2  - m_0^2} = k!\Lx\frac{\gamma_{\chi\psi}}{2}\Rx^k A^{(2k)}\\
\to&(-2\pi\,i)\Lx\frac{\gamma_{\chi\psi}}{2}\Rx^{2k}
    \int\frac{dt\,dy\,dw}{(2\pi)^4}J_{t,y,w}\ (w\,y)^k\,,\\
\int\frac{d^4\ell}{(2\pi)^4}&\frac{
    \mtrxelm{\chi^-}{\feynsl{\ell}}{\psi^-}^k
    \mtrxelm{\psi^-}{\feynsl{\ell}}{\chi^-}^k}
   {\ell^2  - m_0^2} = k!\,\Lx-2\,\gamma_{\chi\psi}\Rx^k A^{(2k)}\\
\to&(-2\pi\,i){\Lx\gamma_{\chi\psi}\Rx^{2k}}
    \int\frac{dt\,dy\,dw}{(2\pi)^4}J_{t,y,w}\ 
    \Lx w\,y - \frac{m_0^2}{\gamma_{\chi\psi}}\Rx^k\,.
\end{split}
\end{equation}
Together, these equations yield the result that
\begin{equation}
\begin{split}
(-2\pi\,i)\int\frac{dt\,dy\,dw}{(2\pi)^4}&J_{t,y,w} = 1\,,\\
(-2\pi\,i)\int\frac{dt\,dy\,dw}{(2\pi)^4}&J_{t,y,w}\LB
   \Lx w\,y - \frac{m_0^2}{\gamma_{\chi\psi}}\Rx^k - (-\,w\,y)^k\RB = 0\,.
\end{split}
\end{equation}
A consistent solution to this set of equations is
\begin{equation}
D(k) = \int\frac{dt\,dy\,dw}{(2\pi)^4}\,J_{t,y,w}\,(w\,y)^k =
  \frac{1}{k+1}\Lx\frac{m_0^2}{\gamma_{\chi\psi}}\Rx^k
\end{equation}

\subsection{Bubble Contributions to the Tadpole}
To obtain the bubble contributions to the tadpole, I add a second
constraint,
\begin{equation}
\delta\Lx(\ell+K_1)^2 - m_1^2\Rx\,.
\end{equation}
This constraint can be imposed by solving 
\begin{equation}
\begin{split}
\label{eq::yorwsoln1}
0=y\,w&\frac{\mtrxelm{\psi^-}{\feynsl{K}_1}{\chi^-}}{t}
    + y\,2\chi\cdot K_1 + w\,2\psi\cdot K_1\\
 &  + t\mtrxelm{\chi^-}{\feynsl{K}_1}{\psi^-}-\frac{m_0^2}{t\,\gamma_{\chi\psi}}
    \mtrxelm{\psi^-}{\feynsl{K}_1}{\chi^-}\\
 &  + S_1 + m_0^2 - m_1^2
\end{split}
\end{equation}
for the values $w=w_0$ or $y=y_0$ and then constructing an integral
table for $t$ and $y$ or $t$ and $w$, respectively.  In fact, one must
average over the two solutions.  The integral tables are constructed
by considering tensor integrals of the form
\begin{equation}
\begin{split}
\int\frac{d^4\ell}{(2\pi)^4}&\left.\frac{(\ell\cdot\psi)^k
    \mtrxelm{\psi^-}{\feynsl{\ell}}{\chi^-}^m}
   {\Lx\ell^2  - m_0^2\Rx\Lx(\ell+K_1)^2-m_1^2\Rx}\right|_{w=w_0}\\
   &= \Lx K_1\cdot\psi\Rx^k\mtrxelm{\psi^-}{\feynsl{K}_1}{\chi^-}^m
    B^{\{1^{k+m}\}}\\
  \to&(-2\pi\,i)^2\Lx\frac{\gamma_{\chi\psi}}{2}\Rx^k(-\gamma_{\chi\psi})^m
     \int\frac{dt\,dy}{(2\pi)^4}J^\prime_{t,y} y^k\,t^m\,,\\
\int\frac{d^4\ell}{(2\pi)^4}&\left.\frac{(\ell\cdot\chi)^k
    \mtrxelm{\psi^-}{\feynsl{\ell}}{\chi^-}^m}
   {\Lx\ell^2  - m_0^2\Rx\Lx(\ell+K_1)^2-m_1^2\Rx}\right|_{y=y_0}\\
   &= \Lx K_1\cdot\chi\Rx^k\mtrxelm{\psi^-}{\feynsl{K}_1}{\chi^-}^m
    B^{\{1^{k+m}\}}\\
  \to&(-2\pi\,i)^2\Lx\frac{\gamma_{\chi\psi}}{2}\Rx^k(-\gamma_{\chi\psi})^m
     \int\frac{dt\,dw}{(2\pi)^4}J^{\prime}_{t,w} w^k\,t^m\,.
\end{split}
\end{equation}
The integral tables are:
\begin{equation}
\begin{split}
\label{eq::tadxbub}
E_w&(k,n-k) = \int\frac{dt\,dy}{(2\pi)^4}J^\prime_{t,y} 
    y^k\,t^{n-k} =\\
&-\frac{(-1)^n}{n+1}\frac{1}{S_1}\Lx\frac{2\psi\cdot K_1}{\gamma_{\chi\psi}}\Rx^k
    \Lx-\frac{\mtrxelm{\psi^-}{\feynsl{K}_1}{\chi^-}}{\gamma_{\chi\psi}}\Rx^{n-k}\\
     &\times\sum_{i=0}^{\lfloor(n-1)/2\rfloor}
   \binom{n-1-i}{i}\Lx\frac{-m_0^2}{S_1}\Rx^i\Lx\frac{S_1+m_0^2-m_1^2}{S_1}
   \Rx^{n-1-2i}\,,\\
E_y&(k,n-k) = \int\frac{dt\,dy}{(2\pi)^4}J^\prime_{t,w} 
    w^k\,t^{n-k} =\\
&-\frac{(-1)^n}{n+1}\frac{1}{S_1}\Lx\frac{2\chi\cdot K_1}{\gamma_{\chi\psi}}\Rx^k
    \Lx-\frac{\mtrxelm{\psi^-}{\feynsl{K}_1}{\chi^-}}{\gamma_{\chi\psi}}\Rx^{n-k}\\
     &\times\sum_{i=0}^{\lfloor(n-1)/2\rfloor}
   \binom{n-1-i}{i}\Lx\frac{-m_0^2}{S_1}\Rx^i\Lx\frac{S_1+m_0^2-m_1^2}{S_1}
   \Rx^{n-1-2i}\,.
\end{split}
\end{equation}

\subsection{Triangle Contributions to the Tadpole}

The triangle contributions to the tadpole come from adding a third constraint,
\begin{equation}
\delta\Lx(\ell+K_2)^2 - m_2^2\Rx\,.
\end{equation}
The extra constraints can be imposed by simultaneously solving
\eqn{eq::yorwsoln1} and
\begin{equation}
\begin{split}
\label{eq::yorwsoln2}
0=y\,w&\frac{\mtrxelm{\psi^-}{\feynsl{K}_2}{\chi^-}}{t}
    + y\,2\chi\cdot K_2 + w\,2\psi\cdot K_2\\
 &  + t\mtrxelm{\chi^-}{\feynsl{K}_2}{\psi^-}-\frac{m_0^2}{t\,\gamma_{\chi\psi}}
    \mtrxelm{\psi^-}{\feynsl{K}_2}{\chi^-}\\
 &  + S_2 + m_0^2 - m_2^2\
\end{split}
\end{equation}
for $y$ and $w$.  There are two solutions, $(w,y) = (w_1,y_1)$ and
$(w,y) = (w_2,y_2)$, which must be averaged over.  This leaves one
free parameter $t$, for which I must derive an integral table.  This
can be done by considering tensor integrals of the form
\begin{equation}
\begin{split}
\int\frac{d^4\ell}{(2\pi)^4}&\frac{
    \mtrxelm{\psi^-}{\feynsl{\ell}}{\chi^-}^n}{\Lx\ell^2  -
    m_0^2\Rx\Lx(\ell+K_1)^2-m_1^2\Rx\Lx(\ell+K_2)^2-m_2^2\Rx}\\
   =\ &\sum_{i=0}^n \mtrxelm{\psi^-}{\feynsl{K}_1}{\chi^-}^{n-i}
     \mtrxelm{\psi^-}{\feynsl{K}_2}{\chi^-}^{i}\,C^{\{1^{n-i}\,2^{i}\}}\\
  \to\ &(-2\pi\,i)^3(-\gamma_{\chi\psi})^n\int\frac{dt}{(2\pi)^4}
    J^{\prime\prime}_t\,t^n\,,
\end{split}
\end{equation}
where $C^{\{1^{n-i}\,2^{i}\}}$ is the $n$-th rank Passarino-Veltman
triangle coefficient whose Lorentz structure has $(n-i)$ powers of $K_1$
and $i$ powers of $K_2$.  I only need the tadpole term from the
Passarino-Veltman coefficients, so my integral table is:
\begin{equation}
\begin{split}
F(n) = \int&\frac{dt}{(2\pi)^4}\ J^{\prime\prime}_t\,t^n\\
    & = \sum_{i=0}^{n}\frac{\mtrxelm{\psi^-}{\feynsl{K}_1}{\chi^-}^{n-i}
          \mtrxelm{\psi^-}{\feynsl{K}_2}{\chi^-}^{i}}{\gamma^n_{\chi\psi}}f_n(i)
\end{split}
\end{equation}
with the $f_n(i)$'s given by
\begin{equation}
\begin{split}
\label{eq::tadxtri}
f_0(0)= &0\,,\\
f_1(0)= & f_1(1) = 0\\
f_2(0)= & \frac{1}{4}\frac{K_1\cdot K_2}{S_1\,\Delta(K_1,K_2)}\,,
   \qquad f_2(1) = -\frac{1}{2}\frac{1}{\Delta(K_1,K_2)}\,,\\
   f_2(2) = & \frac{1}{4}\frac{K_1\cdot K_2}{S_2\,\Delta(K_1,K_2)}\,,\\
   f_3(0) = &\Lx\frac{1}{6}\frac{(S_1+m_0^2-m_1^2)(K_1\cdot
       K_2)}{S_1^2\,\Delta(K_1,K_2)}\right.\\
     &+\frac{1}{8}\frac{S_2+m_0^2-m_2^2}{S_1\,\Delta(K_1,K_2)}
       +\frac{5}{24}\frac{S_2\,(S_2+m_0^2-m_2^2)}{\Delta^2(K_1,K_2)}\\
     &\left.-\frac{5}{24}\frac{S_2\,(S_1+m_0^2-m_1^2)(K_1\cdot
       K_2)}{S_1\,\Delta^2(K_1,K_2)}\Rx\,,\\
   f_3(1) = &\Lx\frac{1}{24}\frac{S_1+m_0^2-m_1^2}{S_1\,\Delta(K_1,K_2)}\right.\\
     &-\frac{5}{24}\frac{(S_2+m_0^2-m_2^2)(K_1\cdot
       K_2)}{\Delta^2(K_1,K_2)}\\
     &\left.+\frac{5}{24}\frac{S_2(S_1+m_0^2-m_1^2)}{\Delta^2(K_1,K_2)}\Rx\,,\\
   f_3(2) = &\left.f_3(1)\right|_{K_1\leftrightarrow K_2\,,\,m_1\leftrightarrow m_2}\,,\\
   f_3(3) = &\left.f_3(0)\right|_{K_1\leftrightarrow K_2\,,\,m_1\leftrightarrow m_2}\,.
\end{split}
\end{equation}

\subsection{The Tadpole Coefficient}
The complete tadpole coefficient is found by combining the pure
tadpole, bubble and triangle contributions.
\begin{equation}
\begin{split}
a_m =\ &
\left.\Inf{w}{\Inf{y}{\Inf{t}{A^{(m)}}}}(t,y,w)\right|_{t\to0\,,(w\,y)^n\to
  D(n)}\\
 -i&\sum_{\rm bubbles}\LB\left.\Inf{y}{\Inf{t}{\left.\tilde{A}_1^{(m)}
      \tilde{A}_2^{(m)}}\right|_{w=w_0}}\right|_{y^k\,t^{m}\to
     E_w(k,m)}\right.\\
   &\hskip20pt + \left.\left.\Inf{w}{\Inf{t}{\left.\tilde{A}_1^{(m)}
      \tilde{A}_2^{(m)}}\right|_{y=y_0}}\right|_{w^k\,t^{m}\to
     E_y(k,m)}\RB\\
 -\frac{1}{2}&\sum_{\rm triangles}\sum_{(w,y)=(w_{1},y_{1})}^{(w_2,y_2)}\left.\Inf{t}{
   \hat{A}^{(m)}_1\,\hat{A}^{(m)}_2\,\hat{A}^{(m)}_3}(t)
   \right|_{t^n\to F(n)}\,,
\end{split}
\end{equation}
where the $\tilde{A}^{(m)}_n$ are the amplitudes formed by cutting a
propagator in $A^{(m)}$ and $\hat{A}^{(m)}_n$ are the amplitudes
formed by cutting two propagators in $A^{(m)}$.

\section{Comments}
The expressions given above assume that all external momenta are
massive and that all internal masses are distinct.  They have been
tested numerically by decomposing tensor integrals (through the fourth
rank tensor box) using Passarino-Veltman techniques and those
described here.  

As long as the external momenta are massive, the only complication in
allowing the internal masses to become degenerate comes from the
tadpoles.  In case of degeneracy, one must be careful not to over
count the contributions from bubbles and triangles.  This can clearly
happen because the starting amplitude for the tadpole cut
$A^{(m)}(\emptyset,\ell)$ is not tied to any external momenta.  Thus,
if one successively adds cuts to the tadpole configuration, one can
reach the same bubble or triangle configuration from different
starting points.  It is safer to construct special cases for
degenerate mass bubbles and triangles and simply sum over the distinct
configurations.

For $S_1\ne0$ and $m_0=m_1\ne0$,
\begin{equation}
\begin{split}
\label{eq::tadxbub0}
E_w&(k,n-k) = \int\frac{dt\,dy}{(2\pi)^4}J^\prime_{t,y} 
    y^k\,t^{n-k} =\\
&\frac{(-1)^n}{n+1}\frac{1}{S_1}\Lx\frac{2\psi\cdot K_1}{\gamma_{\chi\psi}}\Rx^k
    \Lx-\frac{\mtrxelm{\psi^-}{\feynsl{K}_1}{\chi^-}}{\gamma_{\chi\psi}}\Rx^{n-k}\\
     &\times\sum_{i=1}^{\lfloor n/2\rfloor}
   \binom{n-i}{i}\Lx\frac{-m_0^2}{S_1}\Rx^i\,,\\
E_y&(k,n-k) = \int\frac{dt\,dy}{(2\pi)^4}J^\prime_{t,w} 
    w^k\,t^{n-k} =\\
&\frac{(-1)^n}{n+1}\frac{1}{S_1}\Lx\frac{2\chi\cdot K_1}{\gamma_{\chi\psi}}\Rx^k
    \Lx-\frac{\mtrxelm{\psi^-}{\feynsl{K}_1}{\chi^-}}{\gamma_{\chi\psi}}\Rx^{n-k}\\
     &\times\sum_{i=1}^{\lfloor n/2\rfloor}
   \binom{n-i}{i}\Lx\frac{-m_0^2}{S_1}\Rx^i\,.
\end{split}
\end{equation}

Some formul\ae\ must also be modified when external momenta are
massless.
In particular, the bubble integral is no longer an
independent loop-integral function,
\begin{equation}
\begin{split}
\left.B_0(0,m_0,K_1,m_1)\right|_{S_1=0\,,\,m_0\ne m_1} &= 
   \frac{A_0(m_0) - A_0(m_1)}{m_0^2 - m_1^2}\,,\\
\left.B_0(0,m_0,K_1,m_1)\right|_{S_1=0\,,\,m_0 = m_1\ne0} &= 
   (1-\ep)\frac{A_0(m_0)}{m_0^2}\,,\\
\left.B_0(0,m_0,K_1,m_1)\right|_{S_1=0\,,\,m_0 = m_1 = 0} &=0\,.
\end{split}
\end{equation}
In addition to obviating the need to extract the coefficient of the
massless bubbles, this identity alters the extraction of tadpole terms
from bubbles and triangles.  The expressions in \eqn{eq::tadxbub}
become for $S_1=0\,,\,m_0\ne m_1$,
\begin{equation}
\begin{split}
\label{eq::tadxbub1}
E_w(k,n-k) =\frac{(-1)^{k}}{n+1}&\Lx\frac{m_0^2}{m_0^2-m_1^2}\Rx^{n+1}\\
    \times&\frac{(2\psi\cdot K_1)^k\mtrxelm{\psi^-}
    {\feynsl{K}_1}{\chi^-}^{n-k}}{m_0^2\,\gamma^n_{\chi\psi}}\,,\\
E_y(k,n-k) =\frac{(-1)^{k}}{n+1}&\Lx\frac{m_0^2}{m_0^2-m_1^2}\Rx^{n+1}\\
    \times&\frac{(2\chi\cdot K_1)^k\mtrxelm{\psi^-}
    {\feynsl{K}_1}{\chi^-}^{n-k}}{m_0^2\,\gamma^n_{\chi\psi}}\,,
\end{split}
\end{equation}
and for $S_1=0\,,\,m_0= m_1\ne0$,
\begin{equation}
\begin{split}
\label{eq::tadxbub2}
E_w&(k,n-k) = \frac{(-1)^{k}}{n+1}\frac{(2\psi\cdot K_1)^k
    \mtrxelm{\psi^-}{\feynsl{K}_1}{\chi^-}^{n-k}}{m_0^2\,\gamma^n_{\chi\psi}}\,,\\
E_y&(k,n-k) = \frac{(-1)^{k}}{n+1}\frac{(2\chi\cdot K_1)^k
    \mtrxelm{\psi^-}{\feynsl{K}_1}{\chi^-}^{n-k}}{m_0^2\,\gamma^n_{\chi\psi}}\,,
\end{split}
\end{equation}
Clearly this last case is one in which one must be careful to count the
contribution of this bubble only once, since it can be reached two
different starting points.

Formul\ae\ which depend upon Passarino-Veltman triangle coefficients
also change significantly.  Actually, \eqn{eq::bubxtri} is valid in
the $S_2\to0$ limit and does not appear in the $S_1\to0 $ limit since,
the bubble function $\left.B_0(m_0,-K_1,m_1)\right|_{S_1=0}$ is not a
basis integral. \eqn{eq::tadxtri}, however, changes dramatically as
the external momenta vanish and the internal masses become degenerate.
There are five different special cases to consider as one takes
different combinations of $S_n=0$ and $m_n = m_0$ (taking into account
exchange symmetries).  The full set of expressions is given in
\Appendix{sec::specials}.

\section{Conclusions}
I have described a method, based upon generalized unitarity, for
computing the four-dimensional coefficients of scalar loop-integral
functions in one-loop amplitudes in quantum field theory.  The result
is that the coefficients of the loop integrals are determined from the
product of tree-level diagrams that appear at the vertices of the
loop-integral itself.  The method is valid for arbitrary internal and
external masses and can be applied to any one-loop calculation in
quantum field theory.  When combined with one-loop on-shell recursion
relations, this procedure can be used to construct complete one-loop
amplitudes.

One of the features of this method is that it can make use of
the compact representation of tree-level amplitudes found by using
helicity methods.  This formalism, when combined with on-shell
recursion relations to determine the rational terms, as envisioned in
the unitarity bootstrap, shows great promise for constructing an
automated system for generating one-loop amplitudes.
\vfil\eject
\appendix
\section{Special Cases for the Triangle Contribution to the Tadpole}
\label{sec::specials}
\subsection{Case 1: $S_1 = 0$,\, $m_1 \ne m_0$}
\label{app:case1}
Because the massless scalar bubble breaks into a sum of tadpoles, the
special cases gives non-vanishing contributions to $F(1)$.
\begin{equation}
\begin{split}
\label{eq::tadxtri1}
f_0(0)=&0\,,\\
f_1(0)=&
   \frac{1}{2}\frac{1}{(K_1\cdot K_2)(m_0^2-m_1^2)}\,,\qquad f_1(1) = 0\\
   f_2(0) = & -\frac{1}{4}
     \Lx\frac{m_0^2}{(K_1\cdot K_2)(m_0^2-m_1^2)^2}
      + \frac{S_2+m_0^2-m_2^2}{(K_1\cdot K_2)^2(m_0^2-m_1^2)}\right.\\
      &\left.\qquad- \frac{3}{2}\frac{S_2}{(K_1\cdot K_2)^3}\Rx\,,\\
   f_2(1) = &-\frac{3}{8}\frac{1}{(K_1\cdot K_2)^2}\,,\qquad
   f_2(2) =  \frac{1}{4}\frac{1}{S_2\,(K_1\cdot K_2)}\,,\\
   f_3(0) = &\frac{1}{6}\frac{1}{(K_1\cdot K_2)m_0^2}
      + \frac{1}{8}\frac{(S_2+m_0^2-m_2^2)}{m_0^2\,(K_1\cdot K_2)^2}\\
     +& \frac{1}{8}\frac{(S_2+m_0^2-m_2^2)^2}{m_0^2\,(K_1\cdot K_2)^3}
      - \frac{25}{48}\frac{S_2(S_2+m_0^2-m_2^2)}{(K_1\cdot K_2)^4}\\
     +& \frac{5}{16}\frac{S_2^2(m_0^2-m_1^2)}{(K_1\cdot K_2)^5}\,,\\
   f_3(1) = &\frac{5}{16}\frac{S_2\,(m_1^2-m_0^2)}{(K_1\cdot K_2)^4} +
     \frac{1}{3}\frac{S_2+m_0^2-m_2^2}{(K_1\cdot K_2)^3}\\
     &\qquad-\frac{1}{8}\frac{m_0^2}{(m_0^2-m_1^2)(K_1\cdot K_2)^2}\,,\\
   f_3(2) = &-\frac{1}{4}\frac{m_1^2-m_0^2}{(K_1\cdot K_2)^3}
            -\frac{1}{24}\frac{S_2+m_0^2-m_2^2}{S_2\,(K_1\cdot K_2)^2}\,,\\
   f_3(3) = &\frac{1}{8}\frac{(m_1^2-m_0^2)}{S_2\,(K_1\cdot K_2)^2}
     -\frac{1}{6}\frac{S_2+m_0^2-m_2^2}{S_2^2\,(K_1\cdot K_2)}\,.
\end{split}
\end{equation}
\vfil\eject
\subsection{Case 2: $S_1 = 0$,\, $m_1 = m_0$}
\label{app:case2}
This is perhaps the most difficult case and can be tricky, since the
invariant mass of the third external momentum, $K_3=K_1-K_2$ appears
in the denominator.  It can happen that $S_3=0$, in which case one
should be looking at Case 4 (see \Appendix{app:case4}), and if one had
arrived at this configuration by cutting the other leg of the triangle
first, one would naturally have arrived there.  Therefore, the formula
below assumes that $S_3=(K_1-K_2)^2\ne0$.
\begin{equation}
\begin{split}
\label{eq::tadxtri2}
   f_0(0) = &0\,,\\
   f_1(0) = &\frac{1}{2}\frac{1}{m_0^2\,(K_1\cdot K_2)}\,,\\
   f_1(1) = &0\,,\\
   f_2(0) = &-\frac{1}{4}\frac{1}{S_3\,(K_1\cdot K_2)}
      -\frac{1}{4}\frac{(K_1\cdot K_2) + 2\,m_0^2+S_2-m_2^2}{m_0^2\,(K_1\cdot K_2)^2}\,,\\
   f_2(1) = &\frac{1}{4}\frac{1}{S_3\,(K_1\cdot K_2)}\,,\\
   f_2(2) = &\frac{1}{4}\frac{1}{S_2\,(K_1\cdot K_2)}
      -\frac{1}{4}\frac{1}{S_3\,(K_1\cdot K_2)}\,,\\
   f_3(0) = &\frac{1}{2}\frac{7\,S_3-2(m_0^2-m_2^2)}{S_3\,(K_1\cdot K_2)^2}
      +\frac{1}{8}\frac{m_0^2-m_2^2}{S_3\,(K_1\cdot K_2)^2}\\
      &+\frac{1}{24}\frac{4(K_1\cdot K_2)+14\,m_0^2+3\,S_2-3\,m_2^2}
       {m_0^2\,(K_1\cdot K_2)^2}\\
      &+\frac{1}{8}\frac{(S_2-m_2^2)^2}{m_0^2\,(K_1\cdot K_2)^3}
      +\frac{1}{24}\frac{19\,S_2+8\,m_0^2-11\,m_2^2}{(K_1\cdot K_2)^3}\,,\\
   f_3(1) = &-\frac{1}{6}\frac{2\,S_3-m_0^2+m_2^2}{S_3^2\,(K_1\cdot K_2)}
      -\frac{1}{12}\frac{5\,S_3+m_0^2-m_2^2}{S_3\,(K_1\cdot K_2)^2}\,,\\
   f_3(2) = &\frac{1}{12}\frac{S_3-2\,(m_0^2-m_2^2)}{S_3^2\,(K_1\cdot K_2)}\\
      &+\frac{1}{24}\frac{m_0^2-m_2^2}{S_3\,(K_1\cdot K_2)^2}
      -\frac{1}{24}\frac{m_0^2-m_2^2}{S_2\,(K_1\cdot K_2)^2}\,,\\
   f_3(3) = &\frac{1}{6}\frac{S_3+m_0^2-m_2^2}{S_3^2\,(K_1\cdot K_2)}
      -\frac{1}{6}\frac{S_2+m_0^2-m_2^2}{S_2^2\,(K_1\cdot K_2)}\,.
\end{split}
\end{equation}
\vfil\eject
\subsection{Case 3: $S_1 = S_2 = 0$, $m_1 \ne m_0$\,, $m_2 \ne m_0$}
\label{app:case3}
If both $S_1=0$ and $S_2=0$, the third external momentum $K_3=K_1-K_2$
must be massive, $S_3 = -2\,K_1\cdot K_2\ne0$.  In the limit that
$S_3$ becomes small, the diagram describes a one-loop splitting
amplitude, rather than a scattering amplitude.
\begin{equation}
\begin{split}
\label{eq::tadxtri3}
   f_0(0) = &0\,,\\
   f_1(0) = &\frac{1}{2}\frac{1}{(m_0^2-m_1^2)(K_1\cdot K_2)}\,,\\
   f_1(1) = &\frac{1}{2}\frac{1}{(m_0^2-m_2^2)(K_1\cdot K_2)}\,,\\
   f_2(0) = &-\frac{1}{4}\frac{m_0^2}{(m_0^2-m_1^2)^2(K_1\cdot K_2)}
      - \frac{1}{4}\frac{m_0^2-m_2^2}{(m_0^2-m_1^2)(K_1\cdot K_2)^2}\,,\\
   f_2(1) = &-\frac{1}{2}\frac{1}{(K_1\cdot K_2)^2}\,,\\
   f_2(2) = &-\frac{1}{4}\frac{m_0^2}{(m_0^2-m_2^2)^2(K_1\cdot K_2)}
      - \frac{1}{4}\frac{m_0^2-m_1^2}{(m_0^2-m_2^2)(K_1\cdot K_2)^2}\,,\\
   f_3(0) = &\frac{1}{6}\frac{m_0^4}{(m_0^2-m_1^2)^3(K_1\cdot K_2)}
      + \frac{1}{8}\frac{m_0^2(m_0^2-m_2^2)}{(m_0^2-m_1^2)^2(K_1\cdot K_2)^2}\\
      &+ \frac{1}{8}\frac{(m_0^2-m_2^2)^2}{(m_0^2-m_1^2)(K_1\cdot K_2)^3}\,,\\
   f_3(1) = &-\frac{1}{8}\frac{m_0^2}{(m_0^2-m_1^2)(K_1\cdot K_2)^2}
    + \frac{3}{8}\frac{m_0^2 - m_2^2}{(K_1\cdot K_2)^3}\,\\
   f_3(2) = &-\frac{1}{8}\frac{m_0^2}{(m_0^2-m_2^2)(K_1\cdot K_2)^2}
    + \frac{3}{8}\frac{m_0^2 - m_1^2}{(K_1\cdot K_2)^3}\,.\\
   f_3(3) = &\frac{1}{6}\frac{m_0^4}{(m_0^2-m_2^2)^3(K_1\cdot K_2)}
      + \frac{1}{8}\frac{m_0^2(m_0^2-m_1^2)}{(m_0^2-m_2^2)^2(K_1\cdot K_2)^2}\\
      &+ \frac{1}{8}\frac{(m_0^2-m_1^2)^2}{(m_0^2-m_2^2)(K_1\cdot K_2)^3}\,.
\end{split}
\end{equation}
\vfil\eject

\subsection{Case 4: $S_1 = S_2 = 0$,\, $m_1 = m_0$\,, $m_2 \ne
   m_0$}
\label{app:case4}
Again, in a scattering amplitude, $S_3 = -2\,K_1\cdot K_2\ne0$ 
\begin{equation}
\begin{split}
\label{eq::tadxtri4}
   f_0(0) = &0\,,\\
   f_1(0) = &\frac{1}{2}\frac{1}{m_0^2\,(K_1\cdot K_2)}\,,\\
   f_1(1) = &\frac{1}{2}\frac{1}{(m_0^2-m_2^2)(K_1\cdot K_2)}\,,\\
   f_2(0) = &-\frac{1}{4}\frac{(K_1\cdot K_2)-m_2^2}{m_0^2\,(K_1\cdot K_2)^2}
      -\frac{3}{8}\frac{1}{(K_1\cdot K_2)^2}\,,\\
   f_2(1) = &-\frac{1}{4}\frac{1}{(K_1\cdot K_2)^2}\,,\\
   f_2(2) = &-\frac{1}{4}\frac{m_0^2}{(m_0^2-m_2^2)^2(K_1\cdot K_2)}
      + \frac{1}{8}\frac{1}{(K_1\cdot K_2)^2}\,,\\
   f_3(0) = &\frac{1}{6}\frac{1}{m_0^2\,(K_1\cdot K_2)}
      +\frac{7}{24}\frac{1}{(K_1\cdot K_2)^2}\\
      &-\frac{1}{8}\frac{m_2^2\,((K_1\cdot K_2)+m_0^2-m_2^2)}{m_0^2\,(K_1\cdot K_2)^3}
      +\frac{11}{48}\frac{m_0^2-m_2^2}{(K_1\cdot K_2)^3}\,,\\
   f_3(1) = &\frac{1}{8}\frac{2(K_1\cdot K_2)- m_0^2 +
        m_2^2}{(K_1\cdot K_2)^3}\,,\\
   f_3(2) = &-\frac{1}{8}\frac{m_0^2}{(m_0^2-m_2^2)(K_1\cdot K_2)^2}
      -\frac{1}{16}\frac{m_0^2-m_2^2}{(K_1\cdot K_2)^3}\,,\\
   f_3(3) = &\frac{1}{6}\frac{m_0^4}{(m_0^2-m_2^2)^3}
      -\frac{1}{24}\frac{2(K_1\cdot K_2)-m_0^2+m_2^2}{(K_1\cdot K_2)^3}\,,\\
\end{split}
\end{equation}

\subsection{Case 5: $S_1 = S_2 = 0$,\, $m_2 = m_1 = m_0$}
\label{app:case5}
If both $S_1=0$ and $S_2=0$, and all the internal masses are
degenerate, 
\begin{equation}
\begin{split}
\label{eq::tadxtri5}
   f_0(0)=&0\,,\\
   f_1(0)=& f_1(1) = 
   \frac{1}{2}\frac{1}{m_0^2\,(K_1\cdot K_2)}\\
   f_2(0) = & f_2(2) = -\frac{1}{4}\frac{1}{m_0^2\,(K_1\cdot K_2)}\,,\\
   f_2(1) = & 0\,,\\
   f_3(0) = &f_3(3) = \frac{1}{6}\frac{1}{m_0^2\,(K_1\cdot K_2)}
      + \frac{1}{12}\frac{1}{(K_1\cdot K_2)^2}\,,\\
   f_3(1) = &f_3(2) = -\frac{1}{4}\frac{1}{(K_1\cdot K_2)^2}\,.
\end{split}
\end{equation}

\subsection{Case 6: $S_1 = 0$, $S_2,S_3 \ne 0$,\, $m_2 = m_1 = m_0$}
\label{app:case6}
If only $S_1=0$, but all the internal masses are degenerate,
\begin{equation}
\begin{split}
\label{eq::tadxtri6}
   f_0(0)=&0\,,\\
   f_1(0)=& \frac{1}{2}\frac{1}{m_0^2\,(K_1\cdot K_2)}\,,\\
   f_1(1) = &0\,,\\
   f_2(0) = & -\frac{1}{4}\frac{1}{m_0^2\,(K_1\cdot K_2)}
       -\frac{1}{4}\frac{S_2}{m_0^2\,(K_1\cdot K_2)^2}\,,\\
   f_2(1) = & f_2(2) =  0\,,\\
   f_3(0) = &\frac{1}{6}\frac{1}{m_0^2\,(K_1\cdot K_2)}+\frac{1}{6}\frac{1}{S_3\,(K_1\cdot K_2)}\\
      &+ \frac{1}{6}\frac{1}{(K_1\cdot K_2)^2} + \frac{1}{8}\frac{S_2}{m_0^2\,(K_1\cdot K_2)^2}\\
      &+ \frac{1}{3}\frac{S_2}{(K_1\cdot K_2)^3}+ \frac{1}{8}\frac{S_2^2}{m_0^2\,(K_1\cdot K_2)^3}\,,\\
   f_3(1) = &-\frac{1}{6}\frac{1}{S_3\,(K_1\cdot K_2)} -\frac{1}{3}\frac{1}{(K_1\cdot K_2)^2}\,,\\
   f_3(2) = &\frac{1}{6}\frac{1}{S_3\,(K_1\cdot K_2)}\,,\\
   f_3(3) = &\frac{1}{6}\frac{1}{S_2\,(K_1\cdot K_2)}-\frac{1}{6}\frac{1}{S_3\,(K_1\cdot K_2)}\,.
\end{split}
\end{equation}

\subsection{Case 7: $S_1,S_2,S_3 \ne 0$,\, $m_2 = m_1 = m_0$}
\label{app:case7}
If all external masses are non-zero but all the internal masses are degenerate,
\begin{equation}
\begin{split}
\label{eq::tadxtri7}
   f_0(0)=&0\,,\\
   f_1(0)=& f_1(1) = 0\,,\\
   f_2(0) =& f_2(1) = f_2(2) =  0\,,\\
   f_3(0) = &\frac{1}{6}\frac{S_2 - (K_1\cdot K_2)}{S_3\,\Delta(K_1,K_2)}\\
      &+ \frac{1}{6}\frac{S_2}{\Delta(K_1,K_2)\,(K_1\cdot K_2)}
       + \frac{1}{6}\frac{1}{S_1\,(K_1\cdot K_2)}\,,\\
   f_3(1) = &\frac{1}{6}\frac{(K_1\cdot K_2) - S_2}{S_3\,\Delta(K_1,K_2)}\,,\\
   f_3(2) = &\frac{1}{6}\frac{S_2 - S_3 - (K_1\cdot K_2)}{S_3\,\Delta(K_1,K_2)}\,,\\
   f_3(3) = &\frac{1}{6}\frac{S_2\,S_3 - S_2^2 + (S_2+S_3)(K_1\cdot K_2)}{S_2\,S_3\,\Delta(K_1,K_2)}\,.
\end{split}
\end{equation}


\paragraph*{Acknowledgments:}
I would like to thank Zvi Bern for helpful conversations in the early
part of this work.  This research was supported by the
U.~S.~Department of Energy under Contract No.~DE-AC02-98CH10886.
\vskip 100pt



\begin{thebibliography}{68}
\expandafter\ifx\csname natexlab\endcsname\relax\def\natexlab#1{#1}\fi
\expandafter\ifx\csname bibnamefont\endcsname\relax
  \def\bibnamefont#1{#1}\fi
\expandafter\ifx\csname bibfnamefont\endcsname\relax
  \def\bibfnamefont#1{#1}\fi
\expandafter\ifx\csname citenamefont\endcsname\relax
  \def\citenamefont#1{#1}\fi
\expandafter\ifx\csname url\endcsname\relax
  \def\url#1{\texttt{#1}}\fi
\expandafter\ifx\csname urlprefix\endcsname\relax\def\urlprefix{URL }\fi
\providecommand{\bibinfo}[2]{#2}
\providecommand{\eprint}[2][]{\url{#2}}

\bibitem[{\citenamefont{Berends et~al.}(1981)\citenamefont{Berends, Kleiss,
  De~Causmaecker, Gastmans, and Wu}}]{Berends:1981rb}
\bibinfo{author}{\bibfnamefont{F.~A.} \bibnamefont{Berends}},
  \bibinfo{author}{\bibfnamefont{R.}~\bibnamefont{Kleiss}},
  \bibinfo{author}{\bibfnamefont{P.}~\bibnamefont{De~Causmaecker}},
  \bibinfo{author}{\bibfnamefont{R.}~\bibnamefont{Gastmans}}, \bibnamefont{and}
  \bibinfo{author}{\bibfnamefont{T.~T.} \bibnamefont{Wu}},
  \bibinfo{journal}{Phys. Lett.} \textbf{\bibinfo{volume}{B103}},
  \bibinfo{pages}{124} (\bibinfo{year}{1981}).

\bibitem[{\citenamefont{De~Causmaecker
  et~al.}(1982)\citenamefont{De~Causmaecker, Gastmans, Troost, and
  Wu}}]{DeCausmaecker:1981bg}
\bibinfo{author}{\bibfnamefont{P.}~\bibnamefont{De~Causmaecker}},
  \bibinfo{author}{\bibfnamefont{R.}~\bibnamefont{Gastmans}},
  \bibinfo{author}{\bibfnamefont{W.}~\bibnamefont{Troost}}, \bibnamefont{and}
  \bibinfo{author}{\bibfnamefont{T.~T.} \bibnamefont{Wu}},
  \bibinfo{journal}{Nucl. Phys.} \textbf{\bibinfo{volume}{B206}},
  \bibinfo{pages}{53} (\bibinfo{year}{1982}).

\bibitem[{\citenamefont{Xu et~al.}(1984)\citenamefont{Xu, Zhang, and
  Chang}}]{Xu:1984qe}
\bibinfo{author}{\bibfnamefont{Z.}~\bibnamefont{Xu}},
  \bibinfo{author}{\bibfnamefont{D.-H.} \bibnamefont{Zhang}}, \bibnamefont{and}
  \bibinfo{author}{\bibfnamefont{L.}~\bibnamefont{Chang}}
  (\bibinfo{year}{1984}), \bibinfo{note}{tUTP-84/3-TSINGHUA}.

\bibitem[{\citenamefont{Kleiss and Stirling}(1985)}]{Kleiss:1985yh}
\bibinfo{author}{\bibfnamefont{R.}~\bibnamefont{Kleiss}} \bibnamefont{and}
  \bibinfo{author}{\bibfnamefont{W.~J.} \bibnamefont{Stirling}},
  \bibinfo{journal}{Nucl. Phys.} \textbf{\bibinfo{volume}{B262}},
  \bibinfo{pages}{235} (\bibinfo{year}{1985}).

\bibitem[{\citenamefont{Gunion and Kunszt}(1985)}]{Gunion:1985vca}
\bibinfo{author}{\bibfnamefont{J.~F.} \bibnamefont{Gunion}} \bibnamefont{and}
  \bibinfo{author}{\bibfnamefont{Z.}~\bibnamefont{Kunszt}},
  \bibinfo{journal}{Phys. Lett.} \textbf{\bibinfo{volume}{B161}},
  \bibinfo{pages}{333} (\bibinfo{year}{1985}).

\bibitem[{\citenamefont{Hagiwara and Zeppenfeld}(1986)}]{Hagiwara:1985yu}
\bibinfo{author}{\bibfnamefont{K.}~\bibnamefont{Hagiwara}} \bibnamefont{and}
  \bibinfo{author}{\bibfnamefont{D.}~\bibnamefont{Zeppenfeld}},
  \bibinfo{journal}{Nucl. Phys.} \textbf{\bibinfo{volume}{B274}},
  \bibinfo{pages}{1} (\bibinfo{year}{1986}).

\bibitem[{\citenamefont{Xu et~al.}(1987)\citenamefont{Xu, Zhang, and
  Chang}}]{Xu:1986xb}
\bibinfo{author}{\bibfnamefont{Z.}~\bibnamefont{Xu}},
  \bibinfo{author}{\bibfnamefont{D.-H.} \bibnamefont{Zhang}}, \bibnamefont{and}
  \bibinfo{author}{\bibfnamefont{L.}~\bibnamefont{Chang}},
  \bibinfo{journal}{Nucl. Phys.} \textbf{\bibinfo{volume}{B291}},
  \bibinfo{pages}{392} (\bibinfo{year}{1987}).

\bibitem[{\citenamefont{Mangano and Parke}(1991)}]{Mangano:1991by}
\bibinfo{author}{\bibfnamefont{M.~L.} \bibnamefont{Mangano}} \bibnamefont{and}
  \bibinfo{author}{\bibfnamefont{S.~J.} \bibnamefont{Parke}},
  \bibinfo{journal}{Phys. Rept.} \textbf{\bibinfo{volume}{200}},
  \bibinfo{pages}{301} (\bibinfo{year}{1991}).

\bibitem[{\citenamefont{Berends and Giele}(1988)}]{Berends:1987me}
\bibinfo{author}{\bibfnamefont{F.~A.} \bibnamefont{Berends}} \bibnamefont{and}
  \bibinfo{author}{\bibfnamefont{W.~T.} \bibnamefont{Giele}},
  \bibinfo{journal}{Nucl. Phys.} \textbf{\bibinfo{volume}{B306}},
  \bibinfo{pages}{759} (\bibinfo{year}{1988}).

\bibitem[{\citenamefont{Kosower}(1990)}]{Kosower:1989xy}
\bibinfo{author}{\bibfnamefont{D.~A.} \bibnamefont{Kosower}},
  \bibinfo{journal}{Nucl. Phys.} \textbf{\bibinfo{volume}{B335}},
  \bibinfo{pages}{23} (\bibinfo{year}{1990}).

\bibitem[{\citenamefont{Britto et~al.}(2005{\natexlab{a}})\citenamefont{Britto,
  Cachazo, and Feng}}]{Britto:2004ap}
\bibinfo{author}{\bibfnamefont{R.}~\bibnamefont{Britto}},
  \bibinfo{author}{\bibfnamefont{F.}~\bibnamefont{Cachazo}}, \bibnamefont{and}
  \bibinfo{author}{\bibfnamefont{B.}~\bibnamefont{Feng}},
  \bibinfo{journal}{Nucl. Phys.} \textbf{\bibinfo{volume}{B715}},
  \bibinfo{pages}{499} (\bibinfo{year}{2005}{\natexlab{a}}),
  \eprint{hep-th/0412308}.

\bibitem[{\citenamefont{Britto et~al.}(2005{\natexlab{b}})\citenamefont{Britto,
  Cachazo, Feng, and Witten}}]{Britto:2005fq}
\bibinfo{author}{\bibfnamefont{R.}~\bibnamefont{Britto}},
  \bibinfo{author}{\bibfnamefont{F.}~\bibnamefont{Cachazo}},
  \bibinfo{author}{\bibfnamefont{B.}~\bibnamefont{Feng}}, \bibnamefont{and}
  \bibinfo{author}{\bibfnamefont{E.}~\bibnamefont{Witten}},
  \bibinfo{journal}{Phys. Rev. Lett.} \textbf{\bibinfo{volume}{94}},
  \bibinfo{pages}{181602} (\bibinfo{year}{2005}{\natexlab{b}}),
  \eprint{hep-th/0501052}.

\bibitem[{\citenamefont{Bern et~al.}(2005{\natexlab{a}})\citenamefont{Bern,
  Dixon, and Kosower}}]{Bern:2005hs}
\bibinfo{author}{\bibfnamefont{Z.}~\bibnamefont{Bern}},
  \bibinfo{author}{\bibfnamefont{L.~J.} \bibnamefont{Dixon}}, \bibnamefont{and}
  \bibinfo{author}{\bibfnamefont{D.~A.} \bibnamefont{Kosower}},
  \bibinfo{journal}{Phys. Rev.} \textbf{\bibinfo{volume}{D71}},
  \bibinfo{pages}{105013} (\bibinfo{year}{2005}{\natexlab{a}}),
  \eprint{hep-th/0501240}.

\bibitem[{\citenamefont{Bern et~al.}(2005{\natexlab{b}})\citenamefont{Bern,
  Dixon, and Kosower}}]{Bern:2005ji}
\bibinfo{author}{\bibfnamefont{Z.}~\bibnamefont{Bern}},
  \bibinfo{author}{\bibfnamefont{L.~J.} \bibnamefont{Dixon}}, \bibnamefont{and}
  \bibinfo{author}{\bibfnamefont{D.~A.} \bibnamefont{Kosower}},
  \bibinfo{journal}{Phys. Rev.} \textbf{\bibinfo{volume}{D72}},
  \bibinfo{pages}{125003} (\bibinfo{year}{2005}{\natexlab{b}}),
  \eprint{hep-ph/0505055}.

\bibitem[{\citenamefont{Bern et~al.}(1994{\natexlab{a}})\citenamefont{Bern,
  Dixon, Dunbar, and Kosower}}]{Bern:1994zx}
\bibinfo{author}{\bibfnamefont{Z.}~\bibnamefont{Bern}},
  \bibinfo{author}{\bibfnamefont{L.~J.} \bibnamefont{Dixon}},
  \bibinfo{author}{\bibfnamefont{D.~C.} \bibnamefont{Dunbar}},
  \bibnamefont{and} \bibinfo{author}{\bibfnamefont{D.~A.}
  \bibnamefont{Kosower}}, \bibinfo{journal}{Nucl. Phys.}
  \textbf{\bibinfo{volume}{B425}}, \bibinfo{pages}{217}
  (\bibinfo{year}{1994}{\natexlab{a}}), \eprint{hep-ph/9403226}.

\bibitem[{\citenamefont{Bern et~al.}(1995)\citenamefont{Bern, Dixon, Dunbar,
  and Kosower}}]{Bern:1994cg}
\bibinfo{author}{\bibfnamefont{Z.}~\bibnamefont{Bern}},
  \bibinfo{author}{\bibfnamefont{L.~J.} \bibnamefont{Dixon}},
  \bibinfo{author}{\bibfnamefont{D.~C.} \bibnamefont{Dunbar}},
  \bibnamefont{and} \bibinfo{author}{\bibfnamefont{D.~A.}
  \bibnamefont{Kosower}}, \bibinfo{journal}{Nucl. Phys.}
  \textbf{\bibinfo{volume}{B435}}, \bibinfo{pages}{59} (\bibinfo{year}{1995}),
  \eprint{hep-ph/9409265}.

\bibitem[{\citenamefont{Bern and Morgan}(1996)}]{Bern:1995db}
\bibinfo{author}{\bibfnamefont{Z.}~\bibnamefont{Bern}} \bibnamefont{and}
  \bibinfo{author}{\bibfnamefont{A.~G.} \bibnamefont{Morgan}},
  \bibinfo{journal}{Nucl. Phys.} \textbf{\bibinfo{volume}{B467}},
  \bibinfo{pages}{479} (\bibinfo{year}{1996}), \eprint{hep-ph/9511336}.

\bibitem[{\citenamefont{Bern et~al.}(1997)\citenamefont{Bern, Dixon, Dunbar,
  and Kosower}}]{Bern:1996ja}
\bibinfo{author}{\bibfnamefont{Z.}~\bibnamefont{Bern}},
  \bibinfo{author}{\bibfnamefont{L.~J.} \bibnamefont{Dixon}},
  \bibinfo{author}{\bibfnamefont{D.~C.} \bibnamefont{Dunbar}},
  \bibnamefont{and} \bibinfo{author}{\bibfnamefont{D.~A.}
  \bibnamefont{Kosower}}, \bibinfo{journal}{Phys. Lett.}
  \textbf{\bibinfo{volume}{B394}}, \bibinfo{pages}{105} (\bibinfo{year}{1997}),
  \eprint{hep-th/9611127}.

\bibitem[{\citenamefont{Bern et~al.}(1996{\natexlab{a}})\citenamefont{Bern,
  Dixon, and Kosower}}]{Bern:1996je}
\bibinfo{author}{\bibfnamefont{Z.}~\bibnamefont{Bern}},
  \bibinfo{author}{\bibfnamefont{L.~J.} \bibnamefont{Dixon}}, \bibnamefont{and}
  \bibinfo{author}{\bibfnamefont{D.~A.} \bibnamefont{Kosower}},
  \bibinfo{journal}{Ann. Rev. Nucl. Part. Sci.} \textbf{\bibinfo{volume}{46}},
  \bibinfo{pages}{109} (\bibinfo{year}{1996}{\natexlab{a}}),
  \eprint{hep-ph/9602280}.

\bibitem[{\citenamefont{Bern et~al.}(1996{\natexlab{b}})\citenamefont{Bern,
  Dixon, and Kosower}}]{Bern:1996fj}
\bibinfo{author}{\bibfnamefont{Z.}~\bibnamefont{Bern}},
  \bibinfo{author}{\bibfnamefont{L.~J.} \bibnamefont{Dixon}}, \bibnamefont{and}
  \bibinfo{author}{\bibfnamefont{D.~A.} \bibnamefont{Kosower}},
  \bibinfo{journal}{Nucl. Phys. Proc. Suppl.} \textbf{\bibinfo{volume}{51C}},
  \bibinfo{pages}{243} (\bibinfo{year}{1996}{\natexlab{b}}),
  \eprint{hep-ph/9606378}.

\bibitem[{\citenamefont{Bern et~al.}(2000)\citenamefont{Bern, Dixon, and
  Kosower}}]{Bern:2000dn}
\bibinfo{author}{\bibfnamefont{Z.}~\bibnamefont{Bern}},
  \bibinfo{author}{\bibfnamefont{L.~J.} \bibnamefont{Dixon}}, \bibnamefont{and}
  \bibinfo{author}{\bibfnamefont{D.~A.} \bibnamefont{Kosower}},
  \bibinfo{journal}{JHEP} \textbf{\bibinfo{volume}{01}}, \bibinfo{pages}{027}
  (\bibinfo{year}{2000}), \eprint{hep-ph/0001001}.

\bibitem[{\citenamefont{Bern et~al.}(2002)\citenamefont{Bern, De~Freitas, and
  Dixon}}]{Bern:2002tk}
\bibinfo{author}{\bibfnamefont{Z.}~\bibnamefont{Bern}},
  \bibinfo{author}{\bibfnamefont{A.}~\bibnamefont{De~Freitas}},
  \bibnamefont{and} \bibinfo{author}{\bibfnamefont{L.~J.} \bibnamefont{Dixon}},
  \bibinfo{journal}{JHEP} \textbf{\bibinfo{volume}{03}}, \bibinfo{pages}{018}
  (\bibinfo{year}{2002}), \eprint{hep-ph/0201161}.

\bibitem[{\citenamefont{Cachazo
  et~al.}(2004{\natexlab{a}})\citenamefont{Cachazo, Svrcek, and
  Witten}}]{Cachazo:2004kj}
\bibinfo{author}{\bibfnamefont{F.}~\bibnamefont{Cachazo}},
  \bibinfo{author}{\bibfnamefont{P.}~\bibnamefont{Svrcek}}, \bibnamefont{and}
  \bibinfo{author}{\bibfnamefont{E.}~\bibnamefont{Witten}},
  \bibinfo{journal}{JHEP} \textbf{\bibinfo{volume}{09}}, \bibinfo{pages}{006}
  (\bibinfo{year}{2004}{\natexlab{a}}), \eprint{hep-th/0403047}.

\bibitem[{\citenamefont{Zhu}(2004)}]{Zhu:2004kr}
\bibinfo{author}{\bibfnamefont{C.-J.} \bibnamefont{Zhu}},
  \bibinfo{journal}{JHEP} \textbf{\bibinfo{volume}{04}}, \bibinfo{pages}{032}
  (\bibinfo{year}{2004}), \eprint{hep-th/0403115}.

\bibitem[{\citenamefont{Georgiou and Khoze}(2004)}]{Georgiou:2004wu}
\bibinfo{author}{\bibfnamefont{G.}~\bibnamefont{Georgiou}} \bibnamefont{and}
  \bibinfo{author}{\bibfnamefont{V.~V.} \bibnamefont{Khoze}},
  \bibinfo{journal}{JHEP} \textbf{\bibinfo{volume}{05}}, \bibinfo{pages}{070}
  (\bibinfo{year}{2004}), \eprint{hep-th/0404072}.

\bibitem[{\citenamefont{Wu and Zhu}(2004)}]{Wu:2004fba}
\bibinfo{author}{\bibfnamefont{J.-B.} \bibnamefont{Wu}} \bibnamefont{and}
  \bibinfo{author}{\bibfnamefont{C.-J.} \bibnamefont{Zhu}},
  \bibinfo{journal}{JHEP} \textbf{\bibinfo{volume}{07}}, \bibinfo{pages}{032}
  (\bibinfo{year}{2004}), \eprint{hep-th/0406085}.

\bibitem[{\citenamefont{Kosower}(2005)}]{Kosower:2004yz}
\bibinfo{author}{\bibfnamefont{D.~A.} \bibnamefont{Kosower}},
  \bibinfo{journal}{Phys. Rev.} \textbf{\bibinfo{volume}{D71}},
  \bibinfo{pages}{045007} (\bibinfo{year}{2005}), \eprint{hep-th/0406175}.

\bibitem[{\citenamefont{Georgiou et~al.}(2004)\citenamefont{Georgiou, Glover,
  and Khoze}}]{Georgiou:2004by}
\bibinfo{author}{\bibfnamefont{G.}~\bibnamefont{Georgiou}},
  \bibinfo{author}{\bibfnamefont{E.~W.~N.} \bibnamefont{Glover}},
  \bibnamefont{and} \bibinfo{author}{\bibfnamefont{V.~V.} \bibnamefont{Khoze}},
  \bibinfo{journal}{JHEP} \textbf{\bibinfo{volume}{07}}, \bibinfo{pages}{048}
  (\bibinfo{year}{2004}), \eprint{hep-th/0407027}.

\bibitem[{\citenamefont{Abe et~al.}(2005)\citenamefont{Abe, Nair, and
  Park}}]{Abe:2004ep}
\bibinfo{author}{\bibfnamefont{Y.}~\bibnamefont{Abe}},
  \bibinfo{author}{\bibfnamefont{V.~P.} \bibnamefont{Nair}}, \bibnamefont{and}
  \bibinfo{author}{\bibfnamefont{M.-I.} \bibnamefont{Park}},
  \bibinfo{journal}{Phys. Rev.} \textbf{\bibinfo{volume}{D71}},
  \bibinfo{pages}{025002} (\bibinfo{year}{2005}), \eprint{hep-th/0408191}.

\bibitem[{\citenamefont{Dixon et~al.}(2004)\citenamefont{Dixon, Glover, and
  Khoze}}]{Dixon:2004za}
\bibinfo{author}{\bibfnamefont{L.~J.} \bibnamefont{Dixon}},
  \bibinfo{author}{\bibfnamefont{E.~W.~N.} \bibnamefont{Glover}},
  \bibnamefont{and} \bibinfo{author}{\bibfnamefont{V.~V.} \bibnamefont{Khoze}},
  \bibinfo{journal}{JHEP} \textbf{\bibinfo{volume}{12}}, \bibinfo{pages}{015}
  (\bibinfo{year}{2004}), \eprint{hep-th/0411092}.

\bibitem[{\citenamefont{Bern et~al.}(2005{\natexlab{c}})\citenamefont{Bern,
  Forde, Kosower, and Mastrolia}}]{Bern:2004ba}
\bibinfo{author}{\bibfnamefont{Z.}~\bibnamefont{Bern}},
  \bibinfo{author}{\bibfnamefont{D.}~\bibnamefont{Forde}},
  \bibinfo{author}{\bibfnamefont{D.~A.} \bibnamefont{Kosower}},
  \bibnamefont{and}
  \bibinfo{author}{\bibfnamefont{P.}~\bibnamefont{Mastrolia}},
  \bibinfo{journal}{Phys. Rev.} \textbf{\bibinfo{volume}{D72}},
  \bibinfo{pages}{025006} (\bibinfo{year}{2005}{\natexlab{c}}),
  \eprint{hep-ph/0412167}.

\bibitem[{\citenamefont{Birthwright et~al.}(2005)\citenamefont{Birthwright,
  Glover, Khoze, and Marquard}}]{Birthwright:2005ak}
\bibinfo{author}{\bibfnamefont{T.~G.} \bibnamefont{Birthwright}},
  \bibinfo{author}{\bibfnamefont{E.~W.~N.} \bibnamefont{Glover}},
  \bibinfo{author}{\bibfnamefont{V.~V.} \bibnamefont{Khoze}}, \bibnamefont{and}
  \bibinfo{author}{\bibfnamefont{P.}~\bibnamefont{Marquard}},
  \bibinfo{journal}{JHEP} \textbf{\bibinfo{volume}{05}}, \bibinfo{pages}{013}
  (\bibinfo{year}{2005}), \eprint{hep-ph/0503063}.

\bibitem[{\citenamefont{Risager}(2005)}]{Risager:2005vk}
\bibinfo{author}{\bibfnamefont{K.}~\bibnamefont{Risager}},
  \bibinfo{journal}{JHEP} \textbf{\bibinfo{volume}{12}}, \bibinfo{pages}{003}
  (\bibinfo{year}{2005}), \eprint{hep-th/0508206}.

\bibitem[{\citenamefont{Brandhuber
  et~al.}(2005{\natexlab{a}})\citenamefont{Brandhuber, Spence, and
  Travaglini}}]{Brandhuber:2004yw}
\bibinfo{author}{\bibfnamefont{A.}~\bibnamefont{Brandhuber}},
  \bibinfo{author}{\bibfnamefont{B.~J.} \bibnamefont{Spence}},
  \bibnamefont{and}
  \bibinfo{author}{\bibfnamefont{G.}~\bibnamefont{Travaglini}},
  \bibinfo{journal}{Nucl. Phys.} \textbf{\bibinfo{volume}{B706}},
  \bibinfo{pages}{150} (\bibinfo{year}{2005}{\natexlab{a}}),
  \eprint{hep-th/0407214}.

\bibitem[{\citenamefont{Quigley and Rozali}(2005)}]{Quigley:2004pw}
\bibinfo{author}{\bibfnamefont{C.}~\bibnamefont{Quigley}} \bibnamefont{and}
  \bibinfo{author}{\bibfnamefont{M.}~\bibnamefont{Rozali}},
  \bibinfo{journal}{JHEP} \textbf{\bibinfo{volume}{01}}, \bibinfo{pages}{053}
  (\bibinfo{year}{2005}), \eprint{hep-th/0410278}.

\bibitem[{\citenamefont{Bedford
  et~al.}(2005{\natexlab{a}})\citenamefont{Bedford, Brandhuber, Spence, and
  Travaglini}}]{Bedford:2004py}
\bibinfo{author}{\bibfnamefont{J.}~\bibnamefont{Bedford}},
  \bibinfo{author}{\bibfnamefont{A.}~\bibnamefont{Brandhuber}},
  \bibinfo{author}{\bibfnamefont{B.~J.} \bibnamefont{Spence}},
  \bibnamefont{and}
  \bibinfo{author}{\bibfnamefont{G.}~\bibnamefont{Travaglini}},
  \bibinfo{journal}{Nucl. Phys.} \textbf{\bibinfo{volume}{B706}},
  \bibinfo{pages}{100} (\bibinfo{year}{2005}{\natexlab{a}}),
  \eprint{hep-th/0410280}.

\bibitem[{\citenamefont{Bedford
  et~al.}(2005{\natexlab{b}})\citenamefont{Bedford, Brandhuber, Spence, and
  Travaglini}}]{Bedford:2004nh}
\bibinfo{author}{\bibfnamefont{J.}~\bibnamefont{Bedford}},
  \bibinfo{author}{\bibfnamefont{A.}~\bibnamefont{Brandhuber}},
  \bibinfo{author}{\bibfnamefont{B.~J.} \bibnamefont{Spence}},
  \bibnamefont{and}
  \bibinfo{author}{\bibfnamefont{G.}~\bibnamefont{Travaglini}},
  \bibinfo{journal}{Nucl. Phys.} \textbf{\bibinfo{volume}{B712}},
  \bibinfo{pages}{59} (\bibinfo{year}{2005}{\natexlab{b}}),
  \eprint{hep-th/0412108}.

\bibitem[{\citenamefont{Cachazo
  et~al.}(2004{\natexlab{b}})\citenamefont{Cachazo, Svrcek, and
  Witten}}]{Cachazo:2004by}
\bibinfo{author}{\bibfnamefont{F.}~\bibnamefont{Cachazo}},
  \bibinfo{author}{\bibfnamefont{P.}~\bibnamefont{Svrcek}}, \bibnamefont{and}
  \bibinfo{author}{\bibfnamefont{E.}~\bibnamefont{Witten}},
  \bibinfo{journal}{JHEP} \textbf{\bibinfo{volume}{10}}, \bibinfo{pages}{077}
  (\bibinfo{year}{2004}{\natexlab{b}}), \eprint{hep-th/0409245}.

\bibitem[{\citenamefont{Cachazo}(2004)}]{Cachazo:2004dr}
\bibinfo{author}{\bibfnamefont{F.}~\bibnamefont{Cachazo}}
  (\bibinfo{year}{2004}), \eprint{hep-th/0410077}.

\bibitem[{\citenamefont{Britto et~al.}(2005{\natexlab{c}})\citenamefont{Britto,
  Cachazo, and Feng}}]{Britto:2004nj}
\bibinfo{author}{\bibfnamefont{R.}~\bibnamefont{Britto}},
  \bibinfo{author}{\bibfnamefont{F.}~\bibnamefont{Cachazo}}, \bibnamefont{and}
  \bibinfo{author}{\bibfnamefont{B.}~\bibnamefont{Feng}},
  \bibinfo{journal}{Phys. Rev.} \textbf{\bibinfo{volume}{D71}},
  \bibinfo{pages}{025012} (\bibinfo{year}{2005}{\natexlab{c}}),
  \eprint{hep-th/0410179}.

\bibitem[{\citenamefont{Britto et~al.}(2005{\natexlab{d}})\citenamefont{Britto,
  Cachazo, and Feng}}]{Britto:2004nc}
\bibinfo{author}{\bibfnamefont{R.}~\bibnamefont{Britto}},
  \bibinfo{author}{\bibfnamefont{F.}~\bibnamefont{Cachazo}}, \bibnamefont{and}
  \bibinfo{author}{\bibfnamefont{B.}~\bibnamefont{Feng}},
  \bibinfo{journal}{Nucl. Phys.} \textbf{\bibinfo{volume}{B725}},
  \bibinfo{pages}{275} (\bibinfo{year}{2005}{\natexlab{d}}),
  \eprint{hep-th/0412103}.

\bibitem[{\citenamefont{Brandhuber
  et~al.}(2005{\natexlab{b}})\citenamefont{Brandhuber, McNamara, Spence, and
  Travaglini}}]{Brandhuber:2005jw}
\bibinfo{author}{\bibfnamefont{A.}~\bibnamefont{Brandhuber}},
  \bibinfo{author}{\bibfnamefont{S.}~\bibnamefont{McNamara}},
  \bibinfo{author}{\bibfnamefont{B.~J.} \bibnamefont{Spence}},
  \bibnamefont{and}
  \bibinfo{author}{\bibfnamefont{G.}~\bibnamefont{Travaglini}},
  \bibinfo{journal}{JHEP} \textbf{\bibinfo{volume}{10}}, \bibinfo{pages}{011}
  (\bibinfo{year}{2005}{\natexlab{b}}), \eprint{hep-th/0506068}.

\bibitem[{\citenamefont{Quigley and Rozali}(2006)}]{Quigley:2005cu}
\bibinfo{author}{\bibfnamefont{C.}~\bibnamefont{Quigley}} \bibnamefont{and}
  \bibinfo{author}{\bibfnamefont{M.}~\bibnamefont{Rozali}},
  \bibinfo{journal}{JHEP} \textbf{\bibinfo{volume}{03}}, \bibinfo{pages}{004}
  (\bibinfo{year}{2006}), \eprint{hep-ph/0510148}.

\bibitem[{\citenamefont{Britto et~al.}(2005{\natexlab{e}})\citenamefont{Britto,
  Buchbinder, Cachazo, and Feng}}]{Britto:2005ha}
\bibinfo{author}{\bibfnamefont{R.}~\bibnamefont{Britto}},
  \bibinfo{author}{\bibfnamefont{E.}~\bibnamefont{Buchbinder}},
  \bibinfo{author}{\bibfnamefont{F.}~\bibnamefont{Cachazo}}, \bibnamefont{and}
  \bibinfo{author}{\bibfnamefont{B.}~\bibnamefont{Feng}},
  \bibinfo{journal}{Phys. Rev.} \textbf{\bibinfo{volume}{D72}},
  \bibinfo{pages}{065012} (\bibinfo{year}{2005}{\natexlab{e}}),
  \eprint{hep-ph/0503132}.

\bibitem[{\citenamefont{Britto et~al.}(2006)\citenamefont{Britto, Feng, and
  Mastrolia}}]{Britto:2006sj}
\bibinfo{author}{\bibfnamefont{R.}~\bibnamefont{Britto}},
  \bibinfo{author}{\bibfnamefont{B.}~\bibnamefont{Feng}}, \bibnamefont{and}
  \bibinfo{author}{\bibfnamefont{P.}~\bibnamefont{Mastrolia}},
  \bibinfo{journal}{Phys. Rev.} \textbf{\bibinfo{volume}{D73}},
  \bibinfo{pages}{105004} (\bibinfo{year}{2006}), \eprint{hep-ph/0602178}.

\bibitem[{\citenamefont{Mandelstam}(1958)}]{Mandelstam:1958xc}
\bibinfo{author}{\bibfnamefont{S.}~\bibnamefont{Mandelstam}},
  \bibinfo{journal}{Phys. Rev.} \textbf{\bibinfo{volume}{112}},
  \bibinfo{pages}{1344} (\bibinfo{year}{1958}).

\bibitem[{\citenamefont{Landau}(1959)}]{Landau:1959fi}
\bibinfo{author}{\bibfnamefont{L.~D.} \bibnamefont{Landau}},
  \bibinfo{journal}{Nucl. Phys.} \textbf{\bibinfo{volume}{13}},
  \bibinfo{pages}{181} (\bibinfo{year}{1959}).

\bibitem[{\citenamefont{Cutkosky}(1960)}]{Cutkosky:1960sp}
\bibinfo{author}{\bibfnamefont{R.~E.} \bibnamefont{Cutkosky}},
  \bibinfo{journal}{J. Math. Phys.} \textbf{\bibinfo{volume}{1}},
  \bibinfo{pages}{429} (\bibinfo{year}{1960}).

\bibitem[{\citenamefont{Eden et~al.}(1966)\citenamefont{Eden, Landshoff, Olive,
  and Polkinghorne}}]{Eden:1966bk}
\bibinfo{author}{\bibfnamefont{R.~J.} \bibnamefont{Eden}},
  \bibinfo{author}{\bibfnamefont{P.~V.} \bibnamefont{Landshoff}},
  \bibinfo{author}{\bibfnamefont{D.~I.} \bibnamefont{Olive}}, \bibnamefont{and}
  \bibinfo{author}{\bibfnamefont{J.~C.} \bibnamefont{Polkinghorne}},
  \emph{\bibinfo{title}{The Analytic S Matrix}} (\bibinfo{publisher}{Cambridge
  University Press}, \bibinfo{address}{Cambridge, England},
  \bibinfo{year}{1966}).

\bibitem[{\citenamefont{Bern et~al.}(1998)\citenamefont{Bern, Dixon, and
  Kosower}}]{Bern:1997sc}
\bibinfo{author}{\bibfnamefont{Z.}~\bibnamefont{Bern}},
  \bibinfo{author}{\bibfnamefont{L.~J.} \bibnamefont{Dixon}}, \bibnamefont{and}
  \bibinfo{author}{\bibfnamefont{D.~A.} \bibnamefont{Kosower}},
  \bibinfo{journal}{Nucl. Phys.} \textbf{\bibinfo{volume}{B513}},
  \bibinfo{pages}{3} (\bibinfo{year}{1998}), \eprint{hep-ph/9708239}.

\bibitem[{\citenamefont{Bern et~al.}(2004)\citenamefont{Bern, Dixon, and
  Kosower}}]{Bern:2004cz}
\bibinfo{author}{\bibfnamefont{Z.}~\bibnamefont{Bern}},
  \bibinfo{author}{\bibfnamefont{L.~J.} \bibnamefont{Dixon}}, \bibnamefont{and}
  \bibinfo{author}{\bibfnamefont{D.~A.} \bibnamefont{Kosower}},
  \bibinfo{journal}{JHEP} \textbf{\bibinfo{volume}{08}}, \bibinfo{pages}{012}
  (\bibinfo{year}{2004}), \eprint{hep-ph/0404293}.

\bibitem[{\citenamefont{Bern et~al.}(2005{\natexlab{d}})\citenamefont{Bern,
  Del~Duca, Dixon, and Kosower}}]{Bern:2004ky}
\bibinfo{author}{\bibfnamefont{Z.}~\bibnamefont{Bern}},
  \bibinfo{author}{\bibfnamefont{V.}~\bibnamefont{Del~Duca}},
  \bibinfo{author}{\bibfnamefont{L.~J.} \bibnamefont{Dixon}}, \bibnamefont{and}
  \bibinfo{author}{\bibfnamefont{D.~A.} \bibnamefont{Kosower}},
  \bibinfo{journal}{Phys. Rev.} \textbf{\bibinfo{volume}{D71}},
  \bibinfo{pages}{045006} (\bibinfo{year}{2005}{\natexlab{d}}),
  \eprint{hep-th/0410224}.

\bibitem[{\citenamefont{Bern et~al.}(2006)\citenamefont{Bern, Dixon, and
  Kosower}}]{Bern:2005cq}
\bibinfo{author}{\bibfnamefont{Z.}~\bibnamefont{Bern}},
  \bibinfo{author}{\bibfnamefont{L.~J.} \bibnamefont{Dixon}}, \bibnamefont{and}
  \bibinfo{author}{\bibfnamefont{D.~A.} \bibnamefont{Kosower}},
  \bibinfo{journal}{Phys. Rev.} \textbf{\bibinfo{volume}{D73}},
  \bibinfo{pages}{065013} (\bibinfo{year}{2006}), \eprint{hep-ph/0507005}.

\bibitem[{\citenamefont{Berger et~al.}(2006)\citenamefont{Berger, Bern, Dixon,
  Forde, and Kosower}}]{Berger:2006ci}
\bibinfo{author}{\bibfnamefont{C.~F.} \bibnamefont{Berger}},
  \bibinfo{author}{\bibfnamefont{Z.}~\bibnamefont{Bern}},
  \bibinfo{author}{\bibfnamefont{L.~J.} \bibnamefont{Dixon}},
  \bibinfo{author}{\bibfnamefont{D.}~\bibnamefont{Forde}}, \bibnamefont{and}
  \bibinfo{author}{\bibfnamefont{D.~A.} \bibnamefont{Kosower}},
  \bibinfo{journal}{Phys. Rev.} \textbf{\bibinfo{volume}{D74}},
  \bibinfo{pages}{036009} (\bibinfo{year}{2006}), \eprint{hep-ph/0604195}.

\bibitem[{\citenamefont{Forde}(2007)}]{Forde:2007mi}
\bibinfo{author}{\bibfnamefont{D.}~\bibnamefont{Forde}},
  \bibinfo{journal}{Phys. Rev.} \textbf{\bibinfo{volume}{D75}},
  \bibinfo{pages}{125019} (\bibinfo{year}{2007}), \eprint{arXiv:0704.1835
  [hep-ph]}.

\bibitem[{\citenamefont{Dittmaier}(1999)}]{Dittmaier:1998nn}
\bibinfo{author}{\bibfnamefont{S.}~\bibnamefont{Dittmaier}},
  \bibinfo{journal}{Phys. Rev.} \textbf{\bibinfo{volume}{D59}},
  \bibinfo{pages}{016007} (\bibinfo{year}{1999}), \eprint{hep-ph/9805445}.

\bibitem[{\citenamefont{Bern et~al.}(1993)\citenamefont{Bern, Dixon, and
  Kosower}}]{Bern:1992em}
\bibinfo{author}{\bibfnamefont{Z.}~\bibnamefont{Bern}},
  \bibinfo{author}{\bibfnamefont{L.~J.} \bibnamefont{Dixon}}, \bibnamefont{and}
  \bibinfo{author}{\bibfnamefont{D.~A.} \bibnamefont{Kosower}},
  \bibinfo{journal}{Phys. Lett.} \textbf{\bibinfo{volume}{B302}},
  \bibinfo{pages}{299} (\bibinfo{year}{1993}), \eprint{hep-ph/9212308}.

\bibitem[{\citenamefont{Bern et~al.}(1994{\natexlab{b}})\citenamefont{Bern,
  Dixon, and Kosower}}]{Bern:1993kr}
\bibinfo{author}{\bibfnamefont{Z.}~\bibnamefont{Bern}},
  \bibinfo{author}{\bibfnamefont{L.~J.} \bibnamefont{Dixon}}, \bibnamefont{and}
  \bibinfo{author}{\bibfnamefont{D.~A.} \bibnamefont{Kosower}},
  \bibinfo{journal}{Nucl. Phys.} \textbf{\bibinfo{volume}{B412}},
  \bibinfo{pages}{751} (\bibinfo{year}{1994}{\natexlab{b}}),
  \eprint{hep-ph/9306240}.

\bibitem[{\citenamefont{Passarino and Veltman}(1979)}]{Passarino:1978jh}
\bibinfo{author}{\bibfnamefont{G.}~\bibnamefont{Passarino}} \bibnamefont{and}
  \bibinfo{author}{\bibfnamefont{M.~J.~G.} \bibnamefont{Veltman}},
  \bibinfo{journal}{Nucl. Phys.} \textbf{\bibinfo{volume}{B160}},
  \bibinfo{pages}{151} (\bibinfo{year}{1979}).

\bibitem[{\citenamefont{Ossola et~al.}(2007{\natexlab{a}})\citenamefont{Ossola,
  Papadopoulos, and Pittau}}]{Ossola:2006us}
\bibinfo{author}{\bibfnamefont{G.}~\bibnamefont{Ossola}},
  \bibinfo{author}{\bibfnamefont{C.~G.} \bibnamefont{Papadopoulos}},
  \bibnamefont{and} \bibinfo{author}{\bibfnamefont{R.}~\bibnamefont{Pittau}},
  \bibinfo{journal}{Nucl. Phys.} \textbf{\bibinfo{volume}{B763}},
  \bibinfo{pages}{147} (\bibinfo{year}{2007}{\natexlab{a}}),
  \eprint{hep-ph/0609007}.

\bibitem[{\citenamefont{Ossola et~al.}(2007{\natexlab{b}})\citenamefont{Ossola,
  Papadopoulos, and Pittau}}]{Ossola:2007ax}
\bibinfo{author}{\bibfnamefont{G.}~\bibnamefont{Ossola}},
  \bibinfo{author}{\bibfnamefont{C.~G.} \bibnamefont{Papadopoulos}},
  \bibnamefont{and} \bibinfo{author}{\bibfnamefont{R.}~\bibnamefont{Pittau}}
  (\bibinfo{year}{2007}{\natexlab{b}}), \eprint{arXiv:0711.3596 [hep-ph]}.

\bibitem[{\citenamefont{Anastasiou
  et~al.}(2007{\natexlab{a}})\citenamefont{Anastasiou, Britto, Feng, Kunszt,
  and Mastrolia}}]{Anastasiou:2006jv}
\bibinfo{author}{\bibfnamefont{C.}~\bibnamefont{Anastasiou}},
  \bibinfo{author}{\bibfnamefont{R.}~\bibnamefont{Britto}},
  \bibinfo{author}{\bibfnamefont{B.}~\bibnamefont{Feng}},
  \bibinfo{author}{\bibfnamefont{Z.}~\bibnamefont{Kunszt}}, \bibnamefont{and}
  \bibinfo{author}{\bibfnamefont{P.}~\bibnamefont{Mastrolia}},
  \bibinfo{journal}{Phys. Lett.} \textbf{\bibinfo{volume}{B645}},
  \bibinfo{pages}{213} (\bibinfo{year}{2007}{\natexlab{a}}),
  \eprint{hep-ph/0609191}.

\bibitem[{\citenamefont{Anastasiou
  et~al.}(2007{\natexlab{b}})\citenamefont{Anastasiou, Britto, Feng, Kunszt,
  and Mastrolia}}]{Anastasiou:2006gt}
\bibinfo{author}{\bibfnamefont{C.}~\bibnamefont{Anastasiou}},
  \bibinfo{author}{\bibfnamefont{R.}~\bibnamefont{Britto}},
  \bibinfo{author}{\bibfnamefont{B.}~\bibnamefont{Feng}},
  \bibinfo{author}{\bibfnamefont{Z.}~\bibnamefont{Kunszt}}, \bibnamefont{and}
  \bibinfo{author}{\bibfnamefont{P.}~\bibnamefont{Mastrolia}},
  \bibinfo{journal}{JHEP} \textbf{\bibinfo{volume}{03}}, \bibinfo{pages}{111}
  (\bibinfo{year}{2007}{\natexlab{b}}), \eprint{hep-ph/0612277}.

\bibitem[{\citenamefont{Britto and Feng}(2007)}]{Britto:2007tt}
\bibinfo{author}{\bibfnamefont{R.}~\bibnamefont{Britto}} \bibnamefont{and}
  \bibinfo{author}{\bibfnamefont{B.}~\bibnamefont{Feng}}
  (\bibinfo{year}{2007}), \eprint{arXiv:0711.4284 [hep-ph]}.

\bibitem[{\citenamefont{del Aguila and Pittau}(2004)}]{delAguila:2004nf}
\bibinfo{author}{\bibfnamefont{F.}~\bibnamefont{del Aguila}} \bibnamefont{and}
  \bibinfo{author}{\bibfnamefont{R.}~\bibnamefont{Pittau}},
  \bibinfo{journal}{JHEP} \textbf{\bibinfo{volume}{07}}, \bibinfo{pages}{017}
  (\bibinfo{year}{2004}), \eprint{hep-ph/0404120}.

\bibitem[{\citenamefont{Mastrolia}(2007)}]{Mastrolia:2006ki}
\bibinfo{author}{\bibfnamefont{P.}~\bibnamefont{Mastrolia}},
  \bibinfo{journal}{Phys. Lett.} \textbf{\bibinfo{volume}{B644}},
  \bibinfo{pages}{272} (\bibinfo{year}{2007}), \eprint{hep-th/0611091}.

\bibitem[{\citenamefont{Badger et~al.}(2005)\citenamefont{Badger, Glover,
  Khoze, and Svrcek}}]{Badger:2005zh}
\bibinfo{author}{\bibfnamefont{S.~D.} \bibnamefont{Badger}},
  \bibinfo{author}{\bibfnamefont{E.~W.~N.} \bibnamefont{Glover}},
  \bibinfo{author}{\bibfnamefont{V.~V.} \bibnamefont{Khoze}}, \bibnamefont{and}
  \bibinfo{author}{\bibfnamefont{P.}~\bibnamefont{Svrcek}},
  \bibinfo{journal}{JHEP} \textbf{\bibinfo{volume}{07}}, \bibinfo{pages}{025}
  (\bibinfo{year}{2005}), \eprint{hep-th/0504159}.

\bibitem[{\citenamefont{Badger et~al.}(2006)\citenamefont{Badger, Glover, and
  Khoze}}]{Badger:2005jv}
\bibinfo{author}{\bibfnamefont{S.~D.} \bibnamefont{Badger}},
  \bibinfo{author}{\bibfnamefont{E.~W.~N.} \bibnamefont{Glover}},
  \bibnamefont{and} \bibinfo{author}{\bibfnamefont{V.~V.} \bibnamefont{Khoze}},
  \bibinfo{journal}{JHEP} \textbf{\bibinfo{volume}{01}}, \bibinfo{pages}{066}
  (\bibinfo{year}{2006}), \eprint{hep-th/0507161}.

\end{thebibliography}
\end{document}